\documentclass[11pt, a4paper, logo, copyright]{googledeepmind}

\pdftrailerid{redacted}

\makeatletter
\renewcommand\bibentry[1]{\nocite{#1}{\frenchspacing\@nameuse{BR@r@#1\@extra@b@citeb}}}
\makeatother

\usepackage{kantlipsum, lipsum}
\usepackage{dsfont}
\usepackage{gdm-colors}
\usepackage[utf8]{inputenc}   
\usepackage{newunicodechar}   
\usepackage{wasysym,marvosym}
\usepackage{ulem}
\usepackage{hyperref}       
\usepackage{algorithmicx}
\usepackage{algpseudocode}
\usepackage{multirow}
\usepackage{geometry} 
\usepackage[rightcaption]{sidecap} 

\usepackage{tikz}
\usetikzlibrary{shapes.geometric, arrows.meta, positioning}
\usepackage{amsmath}
\usepackage{algorithm}
\usepackage{algpseudocode}
\usepackage{listings}
\usepackage{enumitem} 
\usepackage{lipsum}
\usepackage[most]{tcolorbox}
\definecolor{thinkcolor}{RGB}{227,196,144}
\definecolor{observecolor}{RGB}{153,201,227}
\definecolor{explorecolor}{RGB}{178,217,200}

\newcounter{caseexample}[section]
\renewcommand{\thecaseexample}{\arabic{caseexample}}

\newcounter{promptexample}[section]

\usepackage{array, multirow, tabularx, booktabs, makecell}

\newcommand{\assignmentQuestionName}{Question} 

\usepackage{booktabs}
\usepackage{arydshln}
\usepackage{dashrule}
\usepackage{twemojis}

\usepackage[authoryear, sort&compress, round]{natbib}

\usepackage{bbding}
\usepackage[T1]{fontenc}    
\usepackage{url}            
\usepackage{booktabs}       
\usepackage{nicefrac}       
\usepackage{microtype}      
\usepackage{amsmath}
\usepackage{graphicx}
\usepackage{multicol}
\usepackage[nameinlink]{cleveref}
\usepackage{bbm}
\usepackage{multirow}
\usepackage{soul}
\usepackage{float}
\usepackage{wrapfig}
\usepackage{blindtext}
\usepackage{tablefootnote}
\usepackage{amsfonts}
\usepackage[flushleft]{threeparttable}
\usepackage{colortbl}
\usepackage{mathtools}
\usepackage{bm}
\usepackage{CJKutf8}
\usepackage{makecell}
\usepackage{caption}
\usepackage{capt-of}
\usepackage{array}
\usepackage{calc}      
\usepackage{caption}   
\usepackage{subcaption}  
\usepackage[bottom]{footmisc}
\usepackage{fontawesome}
\usepackage{tabularx}        
\usepackage{siunitx}

\definecolor{hlcolor}{RGB}{206,32,74}
\newcommand{\hlnum}[1]{\textbf{\textcolor{hlcolor}{#1}}}

\renewcommand\tabularxcolumn[1]{m{#1}}
\newcolumntype{C}{>{\centering\arraybackslash}X}
\newcolumntype{L}{>{\raggedright\arraybackslash}X}

\title{DREAM Technical Report}

\author{DREAM Team}

\begin{abstract}
Industrial recommender systems are typically organized as cascaded pipelines of retrieval, ranking, and re-ranking. While efficient, these pipelines suffer from information fragmentation across modules, scattered optimization objectives, rigid rule-based strategies, and weak real-time intent awareness---leaving session-level shifts between browsing, comparison, and purchase largely unaddressed. We present DREAM (\textbf{D}eveloping \textbf{R}ecommender \textbf{E}ngine with \textbf{A}gentic \textbf{M}ethods), an autonomous optimization control architecture that adds a perception-aware, orchestrable, and auditable policy layer atop the existing pipeline rather than replacing it. 
DREAM introduces two core innovations: (1)~a three-tier Intent Engine that fuses on-device signals into structured L0/L1/L2 intent representations with an edge-cloud trigger chain reducing reporting volume to ${\sim}$8.7\%; and (2)~a Meta Engine whose MetaModel performs layered M1$\to$M2$\to$M3 reasoning---intent summarization, strategy planning informed by Strategy Memory, and parameter translation---and dispatches the resulting parameters through a unified outlet with safety guardrails. Both components are continuously optimized by a Reward Dual Loop that couples offline simulation for strategy-space exploration with online feedback for real-outcome calibration, sustaining a continuous cycle of generation, execution, evaluation, and experience accumulation. Large-scale A/B testing on Taobao's homepage feed demonstrates that, with re-ranking control alone, DREAM improves IPV by \hlnum{2.06\%}, Core IPV by \hlnum{2.39\%}, and GMV by \hlnum{0.88\%}; extending control to fine ranking increases these gains to \hlnum{2.71\%}, \hlnum{3.06\%}, and \hlnum{1.31\%}, respectively, while consistently improving PV by more than \hlnum{1\%}. These gains are achieved without replacing any pipeline model or compromising serving stability, validating the viability of agentic meta-control as an industrial recommendation paradigm.
\end{abstract}

\begin{document}
\begin{CJK*}{UTF8}{gkai}

\maketitle

\begin{figure}[htbp]
    \centering
    \includegraphics[width=0.9\linewidth]{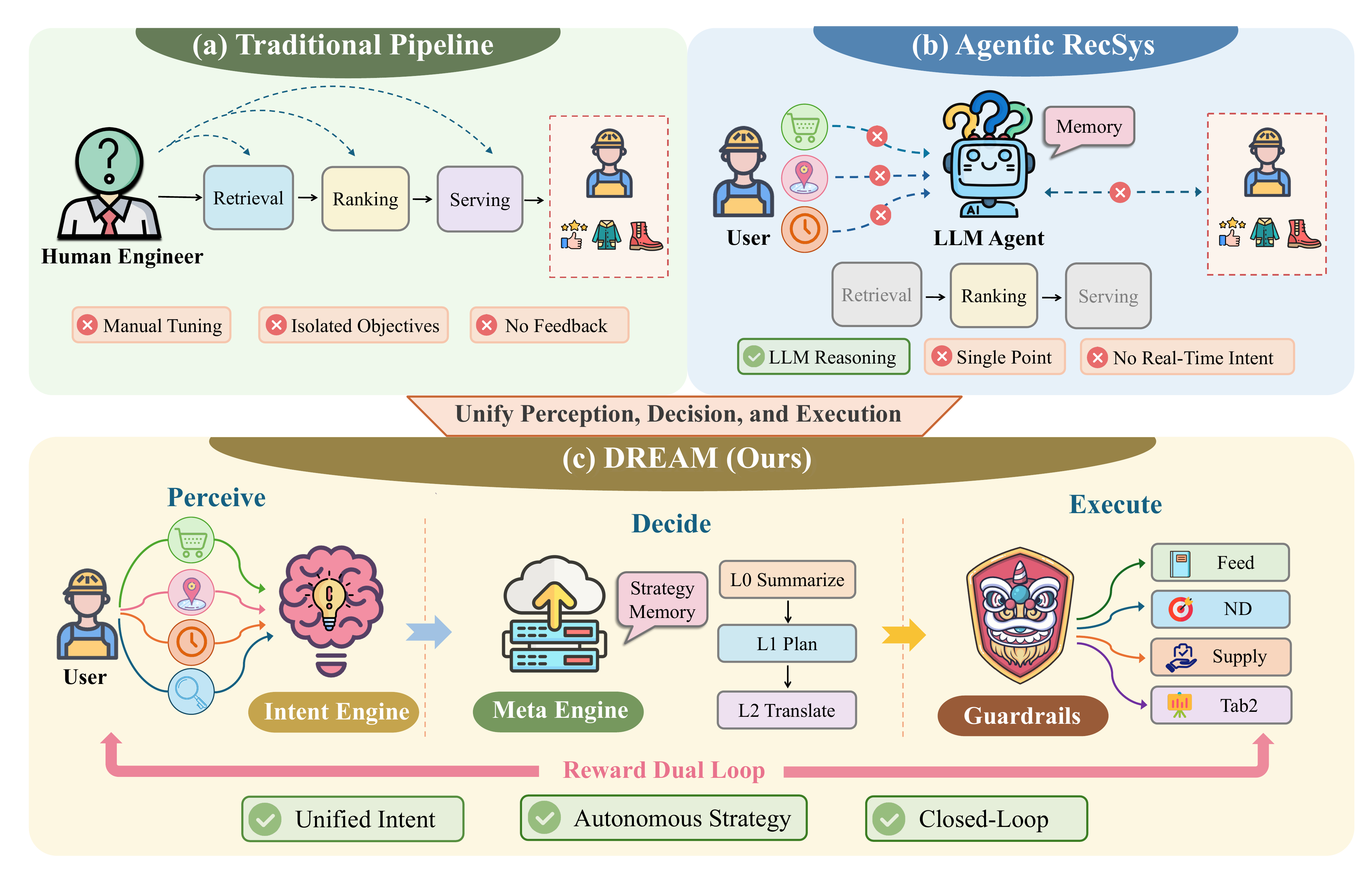}
    \caption{Comparison of recommendation optimization paradigms.}
    \label{fig:dream_teaser}
\end{figure}

\newpage
\setcounter{tocdepth}{3}

\tableofcontents 

\newpage

\section{Introduction}
\label{sec:introduction}

Industrial recommender systems are typically organized as cascaded pipelines of retrieval, ranking, and re-ranking. Such modular designs are stable and efficient, yet they struggle as user behaviors grow more complex and business objectives multiply. Four bottlenecks recur in practice: (1)~\textbf{information fragmentation}, where upstream modules have little visibility into downstream outcomes and vice versa; (2)~\textbf{objective scattering}, with click-through, conversion, growth, and experience optimized independently and never reconciled; (3)~\textbf{strategy rigidity}, since most policies still come from static rules, audience segments, and manual tuning; and (4)~\textbf{weak real-time intent awareness}, leaving session-level shifts between browsing, comparison, and purchase largely unaddressed. The missing piece is a control plane that perceives user intent, coordinates strategies across modules, and self-optimizes from online feedback.

LLM-based agentic recommendation has explored several directions: agents that use memory and tools to select items~\citep{memrec2026,recnet2026,steam2026,amem4rec2026,sager2026,ars2026,chainrec2026,recthinker2026,rrcm2026,twistar2026,reasonrec2026}, agents that simulate users for training and evaluation~\citep{agentcf2023,agentictagger2026,agentgr2026,alignuser2026,abagent2026,hesitator2026}, agents that converse with users~\citep{idss2026,recoworld2025,iagent2025,recbot2025}, and agents that orchestrate the system itself~\citep{selfevolvingrecsys2026,agenticrectune2026,nova2026,automodel2026,evorec2026,agentx2026,sortify2026}. Each tackles one facet of the problem; none spans intent perception, strategy generation, and execution feedback end to end at industrial scale. Three limitations persist across this body of work:

\begin{itemize}
    \item \textbf{Disconnection between intent perception and strategy generation.} Item-selector agents propagate preferences through memory but perceive only individual histories, missing on-device micro-signals, session-level intent, and global business state. Orchestrator agents automate architecture evolution yet plan strategies without real-time user intent, so their policies are far-sighted but poorly aimed.
    \item \textbf{Rigid strategies and weak multi-objective coordination.} Rule-based approaches cannot track real-time state shifts, while monolithic agents fold every objective into one decision with no explicit arbitration. AgenticRecTune~\citep{agenticrectune2026} tunes fusion weights through Pareto Memory, but only at a global level rather than via dynamic per-user policies, slowing iteration and keeping metrics in persistent tension.
    \item \textbf{No closed-loop feedback from execution to upstream optimization.} Orchestrator approaches add harness evolution and A/B judgment, yet their loops serve architecture-level iteration rather than real-time calibration of user-level parameters. Execution signals never flow back to the perception and decision modules, so the optimization cycle is broken.
\end{itemize}


To address these challenges, we propose DREAM (\textbf{D}eveloping \textbf{R}ecommender \textbf{E}ngine with \textbf{A}gentic \textbf{M}ethods), an autonomous optimization control architecture that adds a perception-aware, orchestrable, and auditable policy layer on top of the existing pipeline rather than replacing it. As shown in Figure~\ref{fig:dream_teaser}, DREAM introduces two core innovations:

\textbf{$\bigstar$ Intent-Aware Perception (Intent Engine).}
A three-tier intent engine fuses cross-domain on-device behavioral signals into structured intent representations: L0 (Physical) captures stable profiles, L1 (Demand) captures meta-intent and category needs, and L2 (Preference) captures brand, price, decision stage, and real-time psychology. A cascaded trigger chain escalates only ${\sim}$8.7\% of behavior to cloud-side inference, enabling efficient edge-cloud collaboration. The resulting structured intent, real-time state, and system feedback give the control layer a unified, current view of the user.

\textbf{$\bigstar$ Autonomous Strategy Engine (Meta Engine).}
The Meta Engine folds the control and execution layers into a single strategy loop. It employs a layered reasoning pipeline of M1 intent summarization, M2 strategy planning (augmented by Strategy Memory), and M3 parameter translation. A unified outlet then dispatches the parameters to downstream applications (e.g. \texttt{Homepage Feed}, \texttt{ND}, \texttt{Supply}, and \texttt{TAB2}) using a "default fallback + personalized override" mechanism protected by safety guardrails. Coupling decision and execution this way removes the handoff gap that plagues prior orchestrator designs.

Sustaining both of these components is a \textbf{Reward Dual Loop (Self-Optimizing Feedback)} that closes the optimization cycle independently of either engine's own logic: the offline loop uses an Evaluator to explore the strategy space and refine the LLM through simulation; the online loop measures real user outcomes and deposits conclusions into Strategy Memory. Together they sustain a continuous cycle of "strategy generation $\to$ execution $\to$ evaluation $\to$ experience accumulation $\to$ re-generation", letting the system learn from both simulated and live signal without manual intervention.

DREAM has been validated through large-scale A/B testing on Taobao's homepage feed (Table~\ref{tab:online_main}). With re-ranking control alone, DREAM improves IPV (Item Page Views) by \hlnum{2.06\%}, Core IPV by \hlnum{2.39\%}, and GMV (Gross Merchandise Volume) by \hlnum{0.88\%} without replacing existing pipeline models or compromising serving stability; extending control to fine ranking increases these gains to \hlnum{2.71\%}, \hlnum{3.06\%}, and \hlnum{1.31\%}, respectively, while PV (Page Views) gains consistently exceed \hlnum{1\%}. The cumulative gains from deeper pipeline integration confirm that agentic meta-control benefits compound as the control surface widens.

The system operates through a three-layer pipeline: multi-source user signals are fused into structured intent representations by a three-tier Intent Engine (\S\ref{sec:intent}), which feed into the Meta Engine where MetaModel orchestrates sub-agents for strategy planning and parameter translation (\S\ref{sec:metamodel}). The translated parameters are dispatched through a unified outlet to multiple production scenarios with safety guardrails (\S\ref{sec:dream_frame}). To enable continuous improvement, a Reward Dual Loop couples offline simulation for strategy-space exploration with online feedback for real-outcome calibration, depositing validated conclusions into Strategy Memory for subsequent planning cycles (\S\ref{sec:experiments}).

\section{DREAM Framework}
\label{sec:dream_frame}

\begin{figure}
    \centering
    \captionsetup{justification=centering}
    \includegraphics[width=\textwidth]{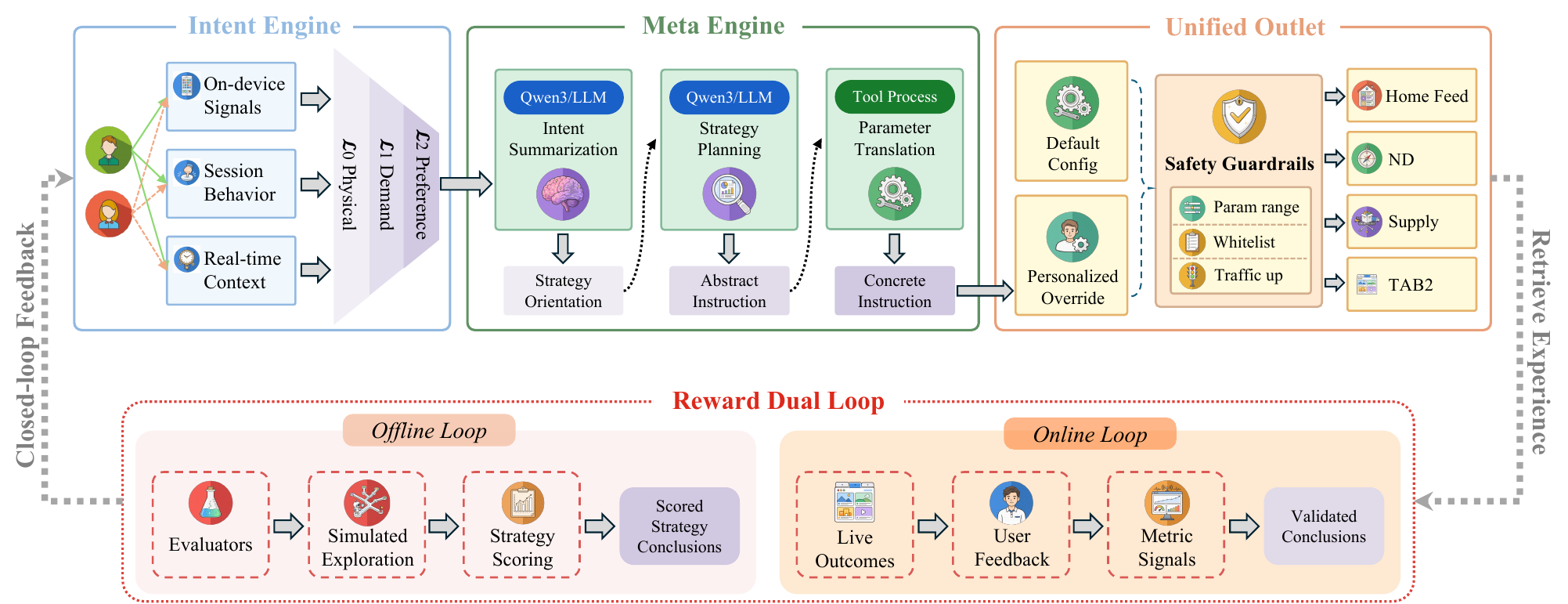}
    \caption{Overview of the DREAM Framework.}
    \label{fig:dream_framework}
\end{figure}

DREAM is an autonomous optimization control architecture that adds, on top of the conventional retrieval-ranking-re-ranking pipeline, a perception-aware and orchestrable policy layer that accumulates experience. Rather than replacing existing stages, DREAM acts as an overlay: it continuously perceives user intent, orchestrates cross-module strategies, and injects personalized execution parameters into the live system through safe override mechanisms.

As illustrated in Figure~\ref{fig:dream_framework}, the framework comprises three core modules that form a collaborative closed loop. The \textbf{Intent Engine} (\S\ref{sec:intent}) forms the perceptual foundation: it turns heterogeneous user behavioral signals into a structured, three-tier intent representation (L0 Physical, L1 Demand, L2 Preference). Designed as an intelligent middleware, it decouples intent production from consumption and exposes one unified representation shared across all downstream modules. The \textbf{Meta Engine} (\S\ref{sec:metamodel}) is the strategic decision layer: a MetaModel (main agent) coordinates specialized sub-agents to diagnose user states, generate candidate strategies, and translate them into bounded executable parameters through a Tools interface. The \textbf{Unified Outlet} injects the orchestrated strategies into downstream scenarios via a personalized override mechanism with a default fallback, while real-time monitoring and fail-safe circuit breaking guarantee deployment safety.

These three modules sustain a closed loop: behavioral signals flow upward from execution to perception, intent representations flow from perception to decision, and strategy parameters flow from decision back to execution. By accumulating intent history and strategy outcomes across sessions, DREAM progressively refines both its intent understanding and its strategy decisions.

\subsection{Intent Engine}
\label{sec:dream_intent}

The Intent Engine is a \textbf{real-time intent recognition and dispatch hub}: an intelligent middleware positioned between upstream behavioral data sources and downstream recommendation applications. It captures, reasons over, and supplies structured user intent at multiple temporal and semantic scales, giving the Meta Engine the grounding it needs for contextually informed strategic decisions.

The engine closes a gap left open by conventional recommender systems: they lack a unified, real-time intent understanding that bridges raw behavioral signals and actionable strategies. Such systems infer intent implicitly, through click-through-rate optimization or sequence modeling, but these approaches offer little semantic interpretability, miss how user needs evolve within and across sessions, and cannot separate co-existing parallel intents. To close this gap, the Intent Engine adopts four design principles: (1)~\textit{decoupled production and consumption}, where the engine optimizes only intent quality while downstream applications decide independently how to consume its output; (2)~\textit{real-time priority}, where explicit behavioral triggers drive intent updates with sub-second latency for high-priority signals; (3)~\textit{explainability}, where every intent output carries a confidence score and an inference rationale; and (4)~\textit{progressive rollout}, where deployment begins with cognitive and heuristic recommendation scenarios before expanding to full coverage.

The engine ingests signals from three orthogonal sources: explicit user actions (searches, clicks, purchases), implicit on-device behaviors (scroll patterns, dwell distributions, cross-product comparisons), and environmental context (temporal, geographic, promotional). It runs a pluggable model architecture coupled with an intelligent routing mechanism that allocates compute according to intent complexity: a 0.8B Main Agent serves the real-time online path, while asynchronous 4B specialized agents refine complex cases and, on nightly idle compute, consolidate the day's behavioral trace into long-term memory. When a request is judged complex, routing escalates it to a stronger context subagent or expert, whose refinements are distilled back into the Main Agent through a self-evolution loop. Inferred intents have the following formal representation:
\begin{equation}
\text{Intent}(u, i, t) = \langle \text{id}, \text{name}, \text{commodity}, \text{strength}, \text{confidence}, \text{origin}, \text{metadata} \rangle
\end{equation}
where each intent carries a strength score, and retained intents are ranked by a priority score $R(i)=s(i)\,\kappa(i)\,\gamma(i)\,u(i)$ that combines behavioral strength $s(i)$, convergence $\kappa(i)$, recency $\gamma(i)$, and unmet demand $u(i)$, normalized before aggregation. To keep online cost bounded, the Main Agent emits only incremental operations (insert/update) instead of rewriting the full intent list, while a harness maintains the cross-turn state keyed by demand category. Alongside this real-time path, a \textit{Dreaming Mechanism} runs a larger 4B model on nightly idle compute to consolidate the day's full-domain behavioral trace into long-term memory through six operations (merge, kill, correct, enrich, add, keep), much as the human brain consolidates episodic experiences during sleep.

Internally, the engine runs a \textbf{trigger-route-evolve-supply} control flow on a device-cloud $F_1$--$F_4$ chain that encodes signals and detects change points on device, then enriches and admits them in the cloud, concentrating computation on moments of genuine intent change while retaining full behavioral coverage. Section~\ref{sec:intent} details the architecture and implementation of the three sub-modules: Multi-Source Data Perception, Intent Reasoning Core, and Downstream Applications.

\subsection{Meta Engine}
\label{sec:dream_metamodel}

The Meta Engine is the strategic decision layer of DREAM. While the Intent Engine determines how the user should be understood at a given moment, the Meta Engine determines what the system should do with that understanding and how to put the resulting decision into effect. It does not replace the existing recall, ranking, and re-ranking pipeline but treats every downstream module as a controllable tool, with structured intent and real-time state as perceptual inputs and the MetaModel as the primary agent for global strategy planning. Below we describe the full decision-to-feedback pipeline: the MetaModel's layered reasoning, Strategy Memory, the Unified Outlet that dispatches parameters, and the Reward Dual Loop that closes the optimization cycle.

The MetaModel is built on Qwen3 and informed by unified perception and historical experience. It converts what the user wants now and what the business needs to optimize into executable strategy instructions, while arbitrating among competing objectives such as conversion versus experience, or exploration versus relevance. It intervenes downstream through a three-stage layered reasoning pipeline: (1)~\textbf{Intent Summarization} condenses the structured intent into a strategy orientation that anchors IPV-oriented or GMV-oriented decision-making; (2)~\textbf{Strategy Planning} produces an abstract instruction (a structured bundle of six strategy modules) while retrieving validated conclusions from Strategy Memory as experiential references; (3)~\textbf{Parameter Translation} maps the abstract instruction into concrete online parameters bounded by predefined schemas and safety constraints. The first two stages run on the nearline path where complex reasoning is affordable; the third runs on the real-time path as lightweight deterministic mapping. All fields take values from strict enumerations, compressing the output space for stability and keeping every strategy decision auditable.

Strategy Memory records whether a strategy is effective for a given class of user state. Indexed by M1 segmentation and updated by the Reward Dual Loop, it supplies positive experiences as candidate references and negative experiences as constraints during Strategy Planning, forming a closed loop from generation to execution, evaluation, consolidation, and re-generation. Full orchestration details are in \S\ref{sec:metamodel}.

The \textbf{Unified Outlet} routes all concrete instructions, whether from the MetaModel or existing online configuration, through a single dispatch path to downstream production scenarios (e.g. \texttt{Homepage Feed}, \texttt{ND}, \texttt{Supply}, and \texttt{TAB2}). Safety guardrails enforce parameter-range bounds, whitelist restrictions, and traffic caps; because every override is incremental and bounded, the mainline pipeline remains the safety net. Details of the controllable parameter space are in \S\ref{sec:metamodel_exec}.

The \textbf{Reward Dual Loop} closes the optimization cycle so that the system continuously self-improves without manual intervention. The Offline Loop uses evaluators to conduct simulated exploration of the strategy space, replaying logged user contexts against candidate bundles to produce scored strategy conclusions. The Online Loop measures live outcomes (user feedback and metric signals such as IPV, CVR, and GMV) against the strategies that produced them, yielding validated conclusions deposited into Strategy Memory. Two feedback paths connect the loop back upstream: Retrieve Experience feeds conclusions into the Meta Engine's Strategy Planning stage, and Closed-loop Feedback propagates execution signals back to the Intent Engine for perception calibration. Together they sustain a continuous cycle of "strategy generation $\to$ execution $\to$ evaluation $\to$ experience accumulation $\to$ re-generation". The offline exploration mechanism is detailed in \S\ref{sec:control_offline_rl}.

\section{Intent Engine}
\label{sec:intent}

\begin{figure}[t]
    \centering
    \includegraphics[width=\linewidth]{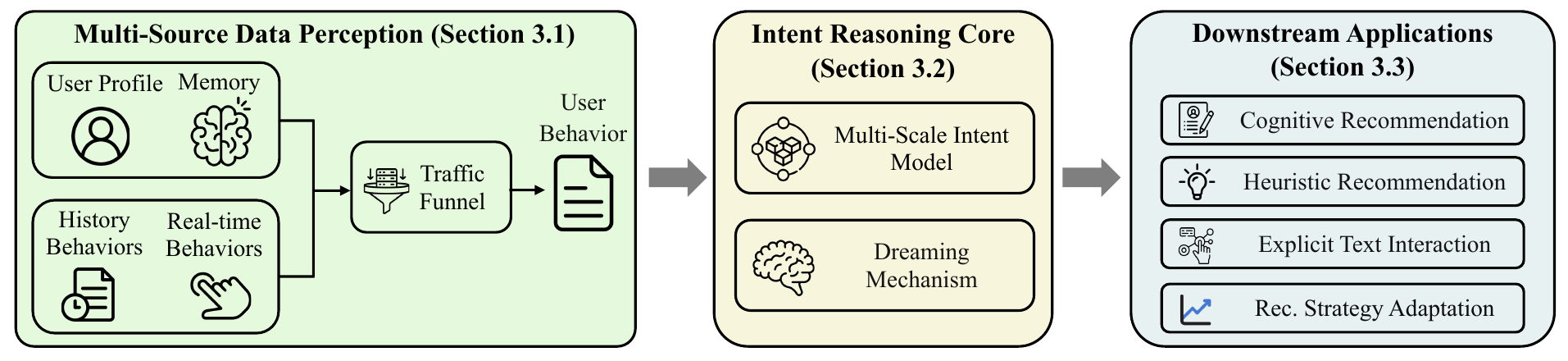}
    \caption{Overview of the Intent Engine: a Multi-Source Data
    Perception layer (Section~\ref{sec:intent_perception}) feeds an
    Intent Reasoning Core (Section~\ref{sec:intent_core}), built on a
    Multi-Scale Intent Model and a Dreaming Mechanism, whose
    hierarchical intent is consumed by Downstream Applications
    (Section~\ref{sec:intent_supply}).}
    \label{fig:intent_overview}
\end{figure}


The Intent Engine is DREAM's foundational perception module: it transforms raw, heterogeneous user signals into a structured, hierarchical intent representation that feeds downstream strategy orchestration directly. Conventional recommender systems infer user intent within isolated pipeline stages, where retrieval models capture coarse interest, ranking models optimize click-through rate, and re-ranking modules apply business rules, yet no unified understanding spans the whole system. The Intent Engine removes this fragmentation by providing a single, coherent perception layer that consolidates multi-source behavioral evidence, reasons over intent at multiple temporal scales, and exposes its output through a standardized hierarchical interface.

A defining trait of the Intent Engine is its design as an intelligent middleware that decouples intent production from consumption: the engine optimizes only intent quality and accuracy, while downstream applications decide independently how to consume the structured output. This separation makes the engine a general-purpose intent infrastructure rather than a component tuned for any single downstream task. 
The architecture of the Intent Engine, illustrated in Figure~\ref{fig:intent_overview}, follows a cascaded \textbf{trigger-route-evolve-supply} control flow comprising three tightly coupled sub-modules:

\begin{itemize}[leftmargin=*,itemsep=2pt]
\item \textbf{Multi-Source Data Perception} (\S\ref{sec:intent_perception}) is the input stage: it ingests signals from explicit user behaviors, implicit on-device behaviors, and environmental context, consolidates the raw stream through a device-cloud $F_1$--$F_4$ chain, and emits a modular structured input for the reasoning core.

\item \textbf{Intent Reasoning Core} (\S\ref{sec:intent_core}) is the processing stage: it converts the enriched signals into structured intent with a 0.8B Main Agent that emits only incremental insert/update operations, a dual-layer router that escalates complex requests to asynchronous 4B specialized agents (a context subagent and an expert), a self-evolution loop that distills their refinements back into the main Agent, and a Dreaming Mechanism (a 4B model) for nightly long-term memory consolidation.

\item \textbf{Downstream Applications} (\S\ref{sec:intent_supply}) is the output stage: it organizes inferred intents into a three-tier representation, comprising the L0 Physical (long-term stable attributes), L1 Demand (intent type, demand category and scene, with confidence), and L2 Preference (fine-grained brand/price/attribute preferences, decision state, and real-time psychological states), and serves them to downstream applications through a top-$K$ selection mechanism governed by intent strength and confidence.
\end{itemize}

The three sub-modules operate as a perception-reasoning-supply pipeline: each module's output is the structured input to the next, while feedback from downstream applications, such as strategy-effectiveness signals from the Meta Engine, flows back to refine perception and reasoning, closing the intent-understanding loop.

\subsection{Multi-Source Data Perception}
\label{sec:intent_perception}

\begin{figure}[t]
    \centering
    \includegraphics[width=\linewidth]{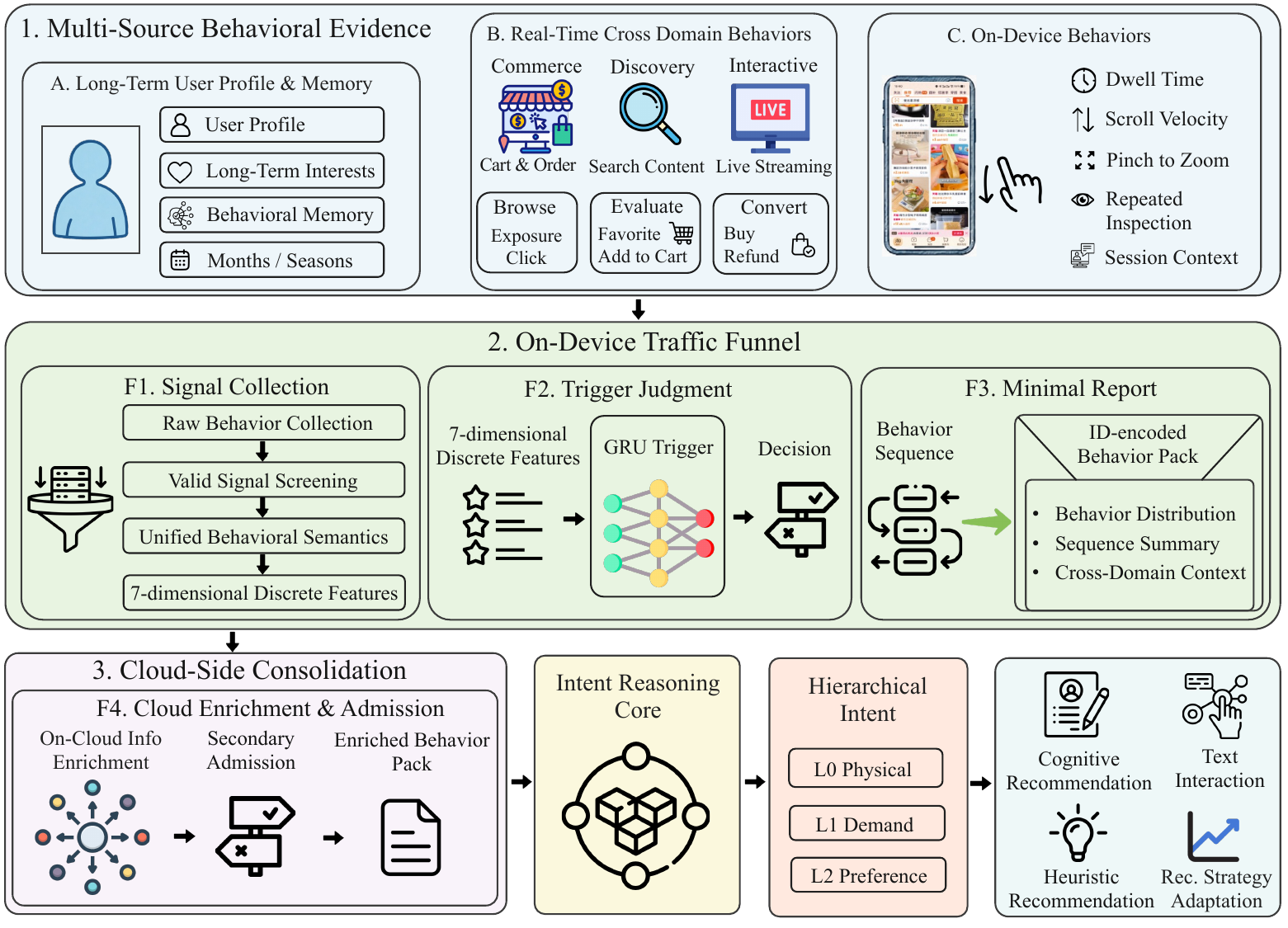}
    \caption{Multi-source data perception in DREAM. A device-cloud
    $F_1$--$F_4$ Traffic Funnel distills multi-source behavioral
    evidence into compact behavior packs: $F_1$--$F_3$ encode, gate, and
    package behaviors on device, while $F_4$ enriches and re-admits them
    in the cloud. The enriched bahavior pack feeds the Intent Reasoning Core to
    maintain the hierarchical L0--L2 intent representation.}
    \label{fig:multi_source_data_perception}
\end{figure}

Accurate intent reasoning is only as good as the behavioral evidence
it consumes.  At Taobao scale, the raw signal space is vast and
heterogeneous: across multiple business domains (item detail, search,
cart, payment, order, live-streaming, content, and beyond), and hundreds of distinct user behaviors are generated
by users, spanning passive exposures,
active clicks, evaluation behaviors (adding items to the cart,
favoriting items, and comparing items), transactions, and explicit
negative feedback (refunds and cart cancellations).
Complementing these real-time interaction streams, two additional
signal sources feed the intent pipeline: (1)~long-lived user profiles
and behavioral memory aggregated over months and seasons, and
(2)~session-level on-device behaviors (page dwell, scroll
velocity, pinch-to-zoom) that capture volatile decision states
invisible to server-side logs.  The Multi-Source Data Perception layer
is responsible for collecting, normalizing and compressing this
heterogeneous signal landscape into a unified, cost-bounded
representation that the downstream Intent Reasoning Core
(Section~\ref{sec:intent_core}) can directly consume.

The central challenge at this layer is a \textbf{signal-acquisition
bottleneck}: the sheer volume of raw behavioral traffic far outstrips
what cloud-side reasoning engines can ingest within latency and budget
constraints, while on-device computational resources, network
bandwidth and power budgets are themselves strictly limited.  A naive
"collect-everything, upload-everything" strategy would saturate the
backbone in seconds; conversely, overly aggressive filtering risks
discarding the very cross-domain, cross-session context on which
high-quality intent inference depends.  The perception layer must
therefore negotiate a careful trade-off between signal
completeness and resource efficiency.

To resolve this tension, we introduce the \textbf{Traffic Funnel}, a
cascaded on-device\,/\,cloud chain ($F_1$--$F_4$) that progressively
filters, compresses and enriches behavioral signals as they propagate
from the client to the cloud. As shown in Figure~\ref{fig:multi_source_data_perception}, the on-device stages distill raw tracking points into a compact structured
encoding of named signals, action IDs and a small set of discrete
feature dimensions, and expand
the usable on-device behavior vocabulary from $6$ to $60{+}$ types.
A lightweight GRU then detects intent-change points so that only about
\hlnum{$15\%$} of behavior is packaged into compact
ID-encoded behavior packs; the cloud side restores full
semantics before a second admission gate decides whether to invoke the
Intent Reasoning Core. This two-level (device and cloud) admission
keeps the reasoning core supplied with the right amount of context at
the right cost.

The Traffic Funnel is the signaling backbone that feeds the Intent
Reasoning Core with the right amount of context at the right cost.
Because a top-of-app recommendation surface routinely observes tens of
thousands of raw user actions per second, dispatching every event to a
cloud-side LLM is feasible in neither latency nor budget. The Traffic
Funnel resolves this by
filtering aggressively at the edge and reasoning selectively in
the cloud, turning a firehose of behavioral events into a small stream
of high-value trigger packets.

We organize the device-cloud pipeline into a strict cascade of four
stages, denoted $F_1$--$F_4$, in which each successive stage runs less
frequently but is more computationally intensive than its predecessor.
The first three stages ($F_1$--$F_3$) live on the client and act as a
triage layer with a tight $<\!100$\,ms budget: they encode raw
tracking points into learnable features, judge when intent is likely
to change, and emit a minimal report. The fourth stage ($F_4$) lives
in the cloud, restoring semantics by ID and applying a second
admission gate before the Intent Reasoning Core is invoked. This
funnel shape, high-frequency and lightweight at the top,
low-frequency and heavyweight at the bottom, explicitly encodes the
end-to-end cost gradient of on-device\,/\,cloud collaboration and
mirrors the $F_1$--$F_4$ topology summarized in
Table~\ref{tab:funnel_stages}. Downstream of $F_4$, intent inference,
the incremental maintenance of the standing intent list, and strategy
generation are handled on the model side (Section~\ref{sec:intent_core}),
decoupling data acquisition from reasoning.

\begin{table}[htbp]
    \centering
    \small
    \renewcommand{\arraystretch}{1.2}
    \begin{tabularx}{\textwidth}{@{}l l l L@{}}
    \toprule
    \textbf{Stage} & \textbf{Locus} & \textbf{Function} & \textbf{Decision output} \\
    \midrule
    $F_1$ & Device & Signal encoding      & raw tracking points encoded into learnable features \\
    $F_2$ & Device & Trigger judgment     & whether an intent-change point fires a report \\
    $F_3$ & Device & Minimal report       & whether an ID-encoded behavior pack is uploaded \\
    $F_4$ & Cloud  & Enrichment + admission & whether the Intent Reasoning Core is invoked \\
    \bottomrule
    \end{tabularx}
    \caption{The four-stage cascaded chain of the Traffic Funnel.
    Each row corresponds to a discrete gating decision; the further
    down the chain a signal travels, the heavier the computation it is
    allowed to trigger.}
    \label{tab:funnel_stages}
\end{table}

\paragraph{On-Device Triage ($F_1$--$F_3$).}
The device-side stages perform the bulk of the compression. They
abstract the raw event vocabulary into a compact, learnable encoding
through a multi-stage pipeline that reduces raw tracking points to
named signals, action IDs, and a small set of discrete feature
dimensions, expanding
the usable on-device behavior vocabulary from $6$ to $60{+}$ types so
that the client can encode a far richer micro-level action trail in
real time.
\begin{itemize}[leftmargin=1.5em, itemsep=2pt, topsep=2pt]
    \item \textbf{$F_1$ Signal Collection.} Rather than waiting for
    T$+$1 batch logs, the client encodes raw tracking points into
    structured behavior features in real time, reducing raw points to
    a compact set of named signals, action IDs and discrete
    feature dimensions. This unified on-device encoding is what
    expands the behavior vocabulary from $6$ to $60{+}$ types and
    preserves the fine-grained action trail for downstream reasoning.
    \item \textbf{$F_2$ Trigger Judgment.} A single-layer GRU performs change-point detection
    over the $7$-dimensional discrete feature stream, carrying its
    hidden state across events so that only likely intent-change
    moments trigger an upload. A short-window mis-tap guard suppresses
    accidental taps (returning to the origin scene within $2$\,s or
    before exposure) to avoid polluting the change-point decision, and
    a RuleTree fallback keeps the gate available whenever the model
    cannot be served. In production this gate admits only about
    \hlnum{$15\%$} of behavior for upload.
    \item \textbf{$F_3$ Minimal Report.} When $F_2$ fires, the device
    packages the recent behavior sequence (at most $50$ events,
    carrying item and shop entity IDs) into an ID-encoded minimal behavior pack and performs no semantic enrichment on device,
    minimizing bandwidth while preserving the full behavior trail.
    Unlike the legacy single-event upload, this pack is a
    session-aggregated payload that delivers the cross-domain context
    on which cloud-side intent reasoning depends.
\end{itemize}
Together, the three device stages turn a high-frequency behavioral
firehose into a sparse stream of high-value reports, with the $F_2$
gate alone admitting only about $15\%$ of behavior for cloud-side
processing.

\paragraph{Cloud-Side Enrichment and Admission ($F_4$).}
Once a minimal behavior pack arrives, the cloud stage $F_4$ restores
the semantics stripped on device and makes a second admission decision.
It joins the pack against the most recent intent list (L1--L2) pulled
from cloud enrichment cache and the L0 static profile (defined as $P_{L0}$), and maps raw identifiers such as
\texttt{item\_id} and \texttt{scene\_id} back to their textual
counterparts (item title, shop, category), at a cost of roughly
$5$\,ms per behavior. A final admission check then decides whether the
enriched update is substantive enough to warrant model-side reasoning,
so that trivial or redundant packs are absorbed at the edge of the
cloud rather than escalated to the Intent Reasoning Core
(Section~\ref{sec:intent_core}).

\noindent
Overall, the Traffic Funnel is a hierarchical gating policy
over the on-device\,/\,cloud cost frontier, where each stage forwards
only a fraction of its incoming traffic to keep every gate both
selective and cheap.

\subsection{Intent Reasoning Core}
\label{sec:intent_core}


The Intent Reasoning Core serves as the cognitive center of the DREAM
Agent system. It
combines behavior packs from the perception layer
(Section~\ref{sec:intent_perception}) with the L0 profile and prior intent
state to produce structured L1 and L2 intent representations. Its design
addresses two requirements: \textbf{timely online inference} within the
serving latency budget and \textbf{consistent intent maintenance} as
behavioral evidence accumulates over time. 

\begin{figure}[!htbp]
    \centering
    \includegraphics[width=0.8\linewidth]{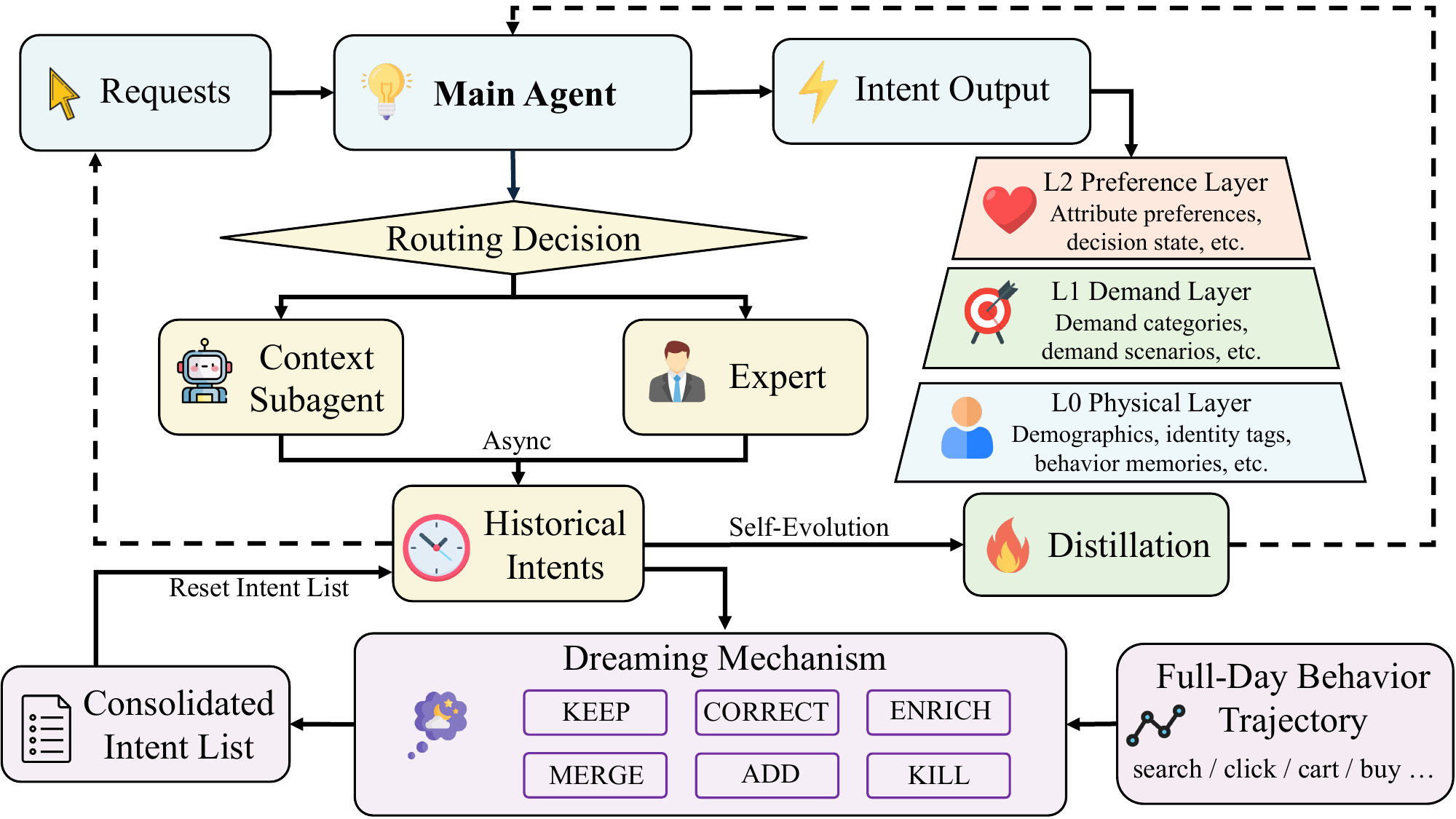}
    \captionsetup{justification=centering}
    \caption{Overall architecture of the Intent Reasoning Core.}
    \label{fig:intent_core}
\end{figure}

These requirements are supported by two complementary mechanisms. The
Multi-Scale Intent Model (Section~\ref{sec:intent_core_model}) performs
online inference through a lightweight Main Agent and selectively invokes
specialized agents for asynchronous refinement and subsequent distillation.
The Dreaming Mechanism
(Section~\ref{sec:intent_core_dreaming}) periodically consolidates the
complete intent state using longer-horizon behavioral evidence. Figure~\ref{fig:intent_core} summarizes the architecture and its information
flow.

\subsubsection{Multi-Scale Intent Model}
\label{sec:intent_core_model}

The Multi-Scale Intent Model consists of five elements: a three-layer intent
schema, a synchronous $0.8$B Main Agent, a dual-layer router, an asynchronous
refinement layer, and a self-evolution loop. The schema defines the structured
output, the Main Agent performs request-time inference, and the router selects
cases that require further processing. The refinement layer produces updated
results for these cases, which are used in subsequent requests and accumulated
for periodic distillation into the Main Agent. 


\paragraph{Hierarchical Intent Schema.}
The Intent Engine represents each user through a shared L0 profile
$P_{L0}$ and a list of active intents. Each active intent contains an L1 demand
representation and an L2 preference representation. L0 provides stable
user context, while L1 and L2 are updated as new behavioral evidence
arrives.

\begin{itemize}[leftmargin=*,itemsep=1pt,topsep=3pt]
    \item \textbf{L0: Physical.}
    L0 contains information shared across all active intents, including
    demographic attributes, persistent interests, long-term behavioral
    memory, and identity tags. It is updated at a relatively coarse
    cadence and serves as a prior for intent reasoning.

    \item \textbf{L1: Demand.}
    L1 describes what the user currently needs. It includes the demand
    category, demand scenario, category cognition, target audience,
    temporal node, and intent confidence. L1 is updated online as new
    behavioral evidence arrives.

    \item \textbf{L2: Preference.}
    L2 describes how the user evaluates alternatives under the
    corresponding L1 demand. It includes subcategory, brand tendency and
    price preference, attribute preference, decision
    state, and real-time psychology. L2 is updated online and is the
    primary target of asynchronous refinement.
\end{itemize}

Each active intent also carries a stable intent ID, a priority from 1 to
5, and one of three intent types: \textit{goal-driven},
\textit{inspiration exploration}, or \textit{aimless browsing}.
Unknown fields are explicitly represented as null values or empty lists
rather than being inferred without sufficient evidence.

\paragraph{Main Agent.}
The Main Agent is a $0.8$B language model that runs synchronously on each
request admitted by the Traffic Funnel. It updates the active intent list
using the L0 profile $P_{L0}$, the previous intent list, and the current behavior
pack. The updated list is returned immediately to downstream applications,
without waiting for asynchronous refinement.

A fixed system prompt specifies the reasoning rules, output schema, and
routing protocol. For each request, the model receives three types of
user-specific context: the L0 profile $P_{L0}$, which provides stable user
characteristics and long-term behavioral information; the historical intent
list, which preserves intent history across requests; and the current
behavior pack, which contains recent interactions and real-time device
signals. The amount of context is adjusted to the serving environment and
its token budget.

Formally, the Main Agent produces an intent update and a routing decision:
\begin{equation}
    \big(\Delta\mathcal{I}_{t}, r_t\big)
    =
    f_{\theta}\big(P_{L0}, \mathcal{I}_{t-1}, \mathcal{B}_{t}\big),
    \label{eq:main_agent}
\end{equation}
where $P_{L0}$ is the L0 profile, $\mathcal{I}_{t-1}$ is the previous
intent list, and $\mathcal{B}_{t}$ is the current behavior pack.
The intent update $\Delta\mathcal{I}_{t}$ contains \textsc{insert} and
\textsc{update} operations. The routing decision $r_t$ specifies whether
the request should be sent to an asynchronous agent and, when applicable,
which agent should process it. Applying the update to the previous list
produces the intent list $\mathcal{I}_{t}$ served for the current request.



\paragraph{Routing.}
After the Main Agent returns, a dual-layer router determines whether the
result requires asynchronous refinement. The rule layer checks the number
of observed behaviors and the number of distinct L1 demand categories. A
request is flagged when either value exceeds its predefined threshold. The
model layer uses the escalation confidence
$c \in [0,1] \cup \{\textsc{null}\}$ that the Main Agent emits as part of
the routing decision $r_t$ in Equation~\ref{eq:main_agent}. A confidence
above the threshold $\tau$ provides an additional escalation signal,
while a \textsc{null} value leaves the decision to the rule layer.

A request is escalated when either layer produces a positive signal. The
router then sends it to the specialized agent selected by the Main Agent.
This process requires no additional model inference on the synchronous path,
and the current intent result is served without waiting for refinement. In
production, approximately $6.3\%$ of requests are sent to the asynchronous
$4$B tier.

\paragraph{Subagents.}
The asynchronous refinement layer contains two specialized $4$B agents
with complementary responsibilities. The \textbf{context subagent} handles
cases in which the L1 demand is plausible but the Subcategory or intent
structure remains uncertain, using longer behavior histories, cross-session
context, and the L0 profile. The \textbf{expert} handles cases in
which the category has been established but item-level preferences remain
uncertain, refining fields such as Price Preference,
Attribute Preference, Decision State, and Intent Type. Selected
requests are processed asynchronously, while the Main Agent output is returned
immediately to the serving pipeline. The refined output is incorporated into
the intent state of subsequent requests. Conflicts are resolved using a
deterministic policy at the field level. Values supported by recent and
reliable behavioral evidence take precedence, while other fields are updated
only when the field-level confidence of the asynchronous estimate exceeds a
predefined threshold.

\paragraph{Self-Evolution.}
\label{sec:intent_core_selfevolve}
Self-evolution operates through intent-state propagation and on-policy
distillation. At the interaction level, the outputs of the context subagent
and the expert trigger an asynchronous update that revises the user's
historical intent list, allowing later
inference to benefit without an immediate parameter update.

At the model level, these outputs are used to train the Main Agent through
on-policy distillation.
Consistent with Equation~\ref{eq:main_agent}, let
$x_t=(P_{L0},\mathcal{I}_{t-1},\mathcal{B}_t)$ denote the request context.
The Main Agent samples an intent-update sequence
$\hat{y}_t\sim p_{\theta}(\cdot\mid x_t)$, where $p_{\theta}$ denotes the
output distribution of the Main Agent $f_{\theta}$. The selected specialized agent
then provides next-token distributions conditioned on the same
student-generated prefixes. The training objective is
\begin{equation}
    \mathcal{L}_{\mathrm{OPD}}(\theta)
    =
    \mathbb{E}_{
        x_t,\,
        \hat{y}_t\sim p_{\theta}(\cdot\mid x_t)
    }
    \left[
        \frac{1}{|\hat{y}_t|}
        \sum_{j=1}^{|\hat{y}_t|}
        \mathrm{KL}
        \left(
            p_{\theta}(\cdot\mid x_t,\hat{y}_{t,<j})
            \,\middle\|\,
            p_{\phi}(\cdot\mid x_t,\hat{y}_{t,<j})
        \right)
    \right],
    \label{eq:opd}
\end{equation}
where $p_{\phi}$ denotes the corresponding distribution of the specialized
$4$B agent. The reverse KL is evaluated along
trajectories generated by the Main Agent, aligning the training distribution
with the states encountered during inference. The objective applies to the
intent-update sequence and does not supervise the routing decision $r_t$.


\paragraph{Evaluation.}
We evaluate the intent model using two protocols based on subsequent
user behavior. In the \textbf{LLM-as-a-Judge} protocol, the judge model
receives the predicted intent output together with the user's subsequent
behavior within the evaluation window. It assesses each structured intent
field according to its consistency with later impressions, searches, and
clicks. Samples with insufficient subsequent behavior are excluded because
they provide limited evidence for evaluating the prediction. The \textbf{Search-Behavior Recall} protocol uses the first search session
observed after intent inference as behavioral evidence. Search term recall
measures whether the predicted intent is semantically consistent with the
user's subsequent query. Filter recall measures whether the predicted
fine-grained preferences correctly anticipate the filtering conditions
selected by the user during search. An LLM-based
evaluator determines semantic matches between the predicted intent
fields and each retained search term or filter.

\begin{center}
    \small
    \renewcommand{\arraystretch}{1.2}
    \begin{tabular}{@{}lccc@{}}
    \toprule
    \textbf{Variant} & \textbf{LLM-as-a-Judge} & \multicolumn{2}{c}{\textbf{Search-Behavior Recall}} \\
    \cmidrule(lr){2-2} \cmidrule(l){3-4}
     & \textbf{Overall Score} & \textbf{Search term} & \textbf{Search Filters} \\
    \midrule
    Baseline                       & 71.32\% & 50.57\% & 24.25\% \\
    \;+ Routing            & 78.20\% & 51.68\% & 26.79\% \\
    \;+ Routing + Self-Evolution   & 84.74\% & 56.05\% & 26.40\% \\
    \bottomrule
    \end{tabular}
    \captionsetup{justification=centering}
    \captionof{table}{Evaluation of the Multi-Scale Intent Model.}
    \label{tab:online_ablation}
\end{center}

As shown in Table~\ref{tab:online_ablation}, routing increases the overall
judge score by $6.88$ percentage points and improves both search term and
filter recall. Adding self-evolution further increases the judge score by
$6.54$ percentage points and search term recall by $4.37$ percentage points.
Filter recall decreases slightly by $0.39$ percentage points but remains
$2.15$ percentage points above the baseline. These results indicate that
the two mechanisms provide complementary gains.

\subsubsection{Dreaming Mechanism}
\label{sec:intent_core_dreaming}

Online inference operates on the compact behavior packs produced by the
Traffic Funnel in Section~\ref{sec:intent_perception}, so each request
reflects only the evidence available at its trigger time. This narrow
horizon leads to three failure modes: \textbf{intent inflation}, where
semantically equivalent intents are created from fragmented evidence;
\textbf{intent inaccuracy}, where fields are inferred before sufficient
evidence has accumulated; and \textbf{intent omission}, where individually
weak but consistent signals never produce an online update. Meanwhile, user
requests drop substantially at night, leaving GPU capacity underutilized.
The Dreaming Mechanism exploits this low-traffic window to consolidate the
intent state without affecting peak-hour serving.

Dreaming applies the same intent
schema to the complete behavior trajectory collected during a day and produces
a consolidated state for subsequent online inference. Let
$\mathcal{I}_{D}$ denote the final online intent list for day $D$, i.e., the
served list $\mathcal{I}_{t}$ after the day's last request, and
$\mathcal{B}_{D}$ the corresponding full-day behavior trajectory. Consistent
with Equation~\ref{eq:main_agent}, a Dreaming pass is defined as
\begin{equation}
    \mathcal{I}_{D+1}^{\mathrm{init}}
    =
    \textsc{Dream}
    \big(
        P_{L0},
        \mathcal{I}_{D},
        \mathcal{B}_{D}
    \big),
    \label{eq:dream}
\end{equation}
where $\mathcal{I}_{D+1}^{\mathrm{init}}$ is the consolidated intent list.
It fully replaces the accumulated online intent state and initializes
$\mathcal{I}_{t-1}$ for the first request of day $D+1$. Unlike the online delta
$\Delta\mathcal{I}_{t}$, which contains only \textsc{insert} and
\textsc{update}, Dreaming operates on complete intent lists using six
operations: \textsc{keep}, \textsc{correct}, \textsc{enrich},
\textsc{merge}, \textsc{add}, and \textsc{kill}.



\paragraph{Triggers and Inputs.}
Dreaming runs as a scheduled daily pass, which is the deployed path and
defines Equation~
\ref{eq:dream}; it may additionally be invoked when the user
remains inactive for a specified interval, when the number of active intents
exceeds a threshold $M$, or when the amount of behavioral evidence not yet
incorporated exceeds a threshold $N$. These auxiliary conditions act as
safeguards: each launches an extra pass over the evidence accumulated so far
using the same operations, without resetting the day index. Every pass
consumes the three inputs of Equation~
ef{eq:dream}: the L0 profile
$P_{L0}$, which supplies stable user context; $\mathcal{I}_{D}$, which holds
the complete intent state accumulated online; and $\mathcal{B}_{D}$, the
temporally ordered full-day behavior trajectory. Relative to a single online
behavior pack, $\mathcal{B}_{D}$ provides broader cross-session evidence,
spanning search, click, cart, and purchase behaviors together with explicit
negative feedback.

\paragraph{Atomic Operations.}
Dreaming transforms the complete intent list using the six operations
summarized in Table~\ref{tab:dream_ops}. Each output records the applied
operation and the affected intent identifiers, making the consolidation
process traceable. These operations act on the complete intent state and
are not restricted to the \textsc{insert} and \textsc{update} operations
used by online inference.

\begin{table}[htbp]
    \centering
    \small
    \renewcommand{\arraystretch}{1.25}
    \begin{tabularx}{\textwidth}{@{}l L l@{}}
    \toprule
    \textbf{Operation} & \textbf{Definition} & \textbf{Function} \\
    \midrule
    \textsc{keep}
    & Preserve an active intent when it remains consistent with the
    extended behavioral evidence.
    & Preservation \\

    \textsc{correct}
    & Revise fields that conflict with the extended behavioral evidence.
    & Error correction \\

    \textsc{enrich}
    & Complete missing or underspecified fields using additional evidence.
    & Information completion \\

    \textsc{merge}
    & Combine multiple entries that refer to the same underlying demand.
    & Deduplication \\

    \textsc{add}
    & Create an intent supported by the extended evidence but absent from
    the online intent list.
    & Omission recovery \\

    \textsc{kill}
    & Remove an intent that has been fulfilled, abandoned, superseded, or
    is no longer supported by current evidence.
    & State retirement \\
    \bottomrule
    \end{tabularx}
    \captionsetup{justification=centering}
    \caption{Atomic operations used by Dreaming.}
    \label{tab:dream_ops}
\end{table}

\paragraph{Consolidation Discipline.}
To maintain stable intent consolidation over long and noisy behavior
sequences, Dreaming follows five evidence-handling rules:

\begin{itemize}[leftmargin=*,itemsep=1pt,topsep=3pt]
    \item \textbf{Outcome-aware reasoning.}
    The model first examines the latest relevant behavior, transaction
    status, and explicit abandonment signals before reviewing earlier
    evidence. This ordering supports reliable \textsc{kill} and
    \textsc{correct} decisions.

    \item \textbf{Behavior clustering and intent alignment.}
    Behaviors are grouped by category, scenario, and time period before
    each group is aligned with an existing intent. This provides coherent
    evidence for \textsc{merge} and \textsc{add}.

    \item \textbf{Evidence-first profile conditioning.}
    The L0 profile is used only when behavioral evidence is insufficient.
    Profile information does not override clear and recent behavioral
    observations.

    \item \textbf{Intent prioritization.}
    Each retained intent $i$ is assigned a priority score
    \[
        R(i)
        =
        s(i)\,\kappa(i)\,\gamma(i)\,u(i),
    \]
    where $s(i)$ denotes behavioral strength, $\kappa(i)$ convergence,
    $\gamma(i)$ recency, and $u(i)$ the degree of unmet demand. The terms
    are normalized before aggregation, and retained intents are ordered
    by decreasing $R(i)$.

    \item \textbf{Attribute-level negative evidence.}
    Explicit negative behaviors are mapped to attribute-level exclusions
    when supported by the surrounding evidence. They are not generalized
    to category-level exclusions, which could remove relevant candidates
    from downstream recall.
\end{itemize}

\paragraph{Evaluation.}
We evaluate Dreaming on a matched set of $683$ users. For each user,
the final online intent list and its consolidated counterpart are evaluated
against the same next-day behavioral evidence using the LLM-as-a-Judge
protocol. The paired design reduces variation caused by differences in user
behavior and allows a direct comparison between the two intent states.


\begin{table}[htbp]
    \centering
    \small
    \renewcommand{\arraystretch}{1.2}
    \begin{tabular}{@{}llrr@{}}
    \toprule
    \textbf{Layer} & \textbf{Field} & \textbf{Daytime} & \textbf{After Dreaming} \\
    \midrule
    \multirow{2}{*}{Intent}
        & Intent type        & $0.583$ & $\mathbf{0.680}$ \\
        & Priority           & $0.524$ & $\mathbf{0.571}$ \\
    \midrule
    \multirow{5}{*}{L1}
        & Demand category    & $0.570$ & $\mathbf{0.580}$ \\
        & Category cognition & $0.819$ & $\mathbf{0.838}$ \\
        & Demand scenario   & $0.797$ & $\mathbf{0.817}$ \\
        & Temporal node      & $0.893$ & $\mathbf{0.897}$ \\
        & Intent confidence  & $0.592$ & $\mathbf{0.611}$ \\
    \midrule
    \multirow{2}{*}{L2}
        & Subcategory   & $0.477$ & $\mathbf{0.523}$ \\
        & Decision state     & $0.561$ & $\mathbf{0.592}$ \\
    \bottomrule
    \end{tabular}
    \captionsetup{justification=centering}
    \caption{Dreaming evaluation.}
    \label{tab:dream_effect}
\end{table}


As shown in Table~\ref{tab:dream_effect}, Dreaming improves all evaluated
fields. The largest gains appear in Intent Type, Priority, and Subcategory, indicating stronger intent organization and category resolution,
while fields that are already reliable online, such as temporal node and
Need Category, improve only marginally.
\subsection{Downstream Applications}
\label{sec:intent_supply}

The Intent Engine creates value when its three-level structured output is consumed by downstream applications. Table~\ref{tab:intent_application} summarizes the output of the Intent Engine  consumed by each downstream scenario. Based on how intent signals are used in the recommendation pipeline, we group downstream applications into three categories: \textbf{recall}, which converts intent signals into candidate item sets; \textbf{explicit text interaction}, which turns intent signals into user-facing natural-language copy; and \textbf{recommendation strategy adaptation}, which translates intent signals into dynamic runtime parameters for the recommendation engine. The quantitative results of these deployments are reported in Section ~\ref{sec:intent_downstream_ab}.

\begin{table}[htbp]
  \centering
  \small
  \renewcommand{\arraystretch}{1.2}
  \renewcommand\tabularxcolumn[1]{m{#1}}
  \caption{The output of Intent Engine consumed by downstream application scenarios.}
  \label{tab:intent_application}
  \begin{tabularx}{\textwidth}{@{} l >{\raggedright\arraybackslash}X l @{}}
    \toprule
    \textbf{Application Scenario} & \textbf{Intent Fields Consumed} & \textbf{Implementation} \\
    \midrule
    \textit{Recall} & & \\[-1pt]
    \quad Cognitive Recommendation & L1 Demand Category; L2 Subcategory & Tag-based Recall \\
    \addlinespace[2pt]
    \quad Heuristic Recommendation & L1 Demand Category; L2 Subcategory; L2~Attribute Preference--Priority Attribute & Inquiry Card \\
    \midrule
    Explicit Text Interaction & L0 User Profile; L1 Demand Category \&~Demand Scenario; L2 Decision State & Inquiry Card \\
    \midrule
    Recommendation Strategy Adaptation & L1 Demand Category \&~L1 Category Cognition; L2 Subcategory \&~Priority & Personalized Parameter \\
    \bottomrule
  \end{tabularx}
\end{table}

\paragraph{Recall.} The structured output of the Intent Engine is organized into three levels of granularity, each providing signals for recall. In L0, the basic user profile and long-term interests form the user prior. In L1, demand category and demand scenario define the recall scope. In L2, subcategory and attribute preference add fine-grained constraints. Intent-driven recall has been deployed in two scenarios: Cognitive Recommendation and Heuristic Recommendation, which consume these signals in different forms: the former uses intent tags as input to the tag-based recall model, while the latter converts intent into Inquiry Card that surfaces user needs directly.

Cognitive Recommendation retrieves items using user-level item tags. Previously, these tags were generated from historical user behavior by specialized models. After the Intent Engine was integrated, L1 demand category tags and L2 subcategory tags were used as a unified input to the tag-based recall model. These intent-derived tags complement the existing model-generated tags, providing explicit and hierarchically structured need semantics for tag-based recall.

Heuristic Recommendation surfaces potential user needs through Inquiry Card. The Inquiry Card is an interactive card shown in the recommendation feed alongside ordinary item cards. It displays a text snippet that articulates the user's need; tapping it leads to a landing page with items matched to the need. After integrating Intent Engine, the system draws on intent signals at different levels: L1 demand category, L2 subcategory and L2 attribute preference-priority attribute. Together, these signals cover user states ranging from open-ended exploration to clearly defined goals.

\paragraph{Explicit Text Interaction.}  Explicit text interaction determines how to surface and communicate the underlying need in natural language. Its primary output is the personalized copy shown on Inquiry Card in Heuristic Recommendation. 

Hierarchical intent fields contribute three types of information. L0 user profile and contextual signals provide personal and situational context; L1 demand category and demand scenario determine the topic and angle of the copy; L2 decision state determines the messaging strategy. These signals are combined with a range of copy styles informed by consumer psychology to produce context-appropriate, differentiated copy. 

\paragraph{Recommendation Strategy Adaptation.} Recommendation Strategy Adaptation dynamically adjusts engine parameters according to the user's current intent state. Users' expectations of recommendation results differ fundamentally between intent states. The differences cannot be satisfied by fixed parameters and must instead be adjusted in real time based on the intent state. For users with concentrated needs, the relevant channels are expanded in scale and coverage to strengthen candidate retrieval of target content; for users with dispersed intent, some channels are scaled down to reduce noisy candidates.

The output of the Intent Engine is fed into a lightweight LLM reasoning module for inference, which generates personalized parameters. The recall channel switches are used to determine the type of recall, the quota parameters control the number of candidates output by each channel, and diversification parameters determine the maximum number of candidates of each category.

The three-level output of Intent Engine is consumed across recall, interaction, and strategy adaptation. The recommendations then influence subsequent user behavior, which drives the next Intent Engine output update. 

Within this closed loop, the Intent Engine is responsible for intent perception. The Meta Engine determines when reasoning is triggered, how often it runs, and how recommendation strategies are orchestrated. These mechanisms are described in detail in the next section.
\section{Meta Engine}
\label{sec:metamodel}


\subsection{Agent Loop Trigger and Frequency Control}
\label{sec:intent_frequency_control}

\begin{figure}[t]
    \centering
    \includegraphics[width=\textwidth]{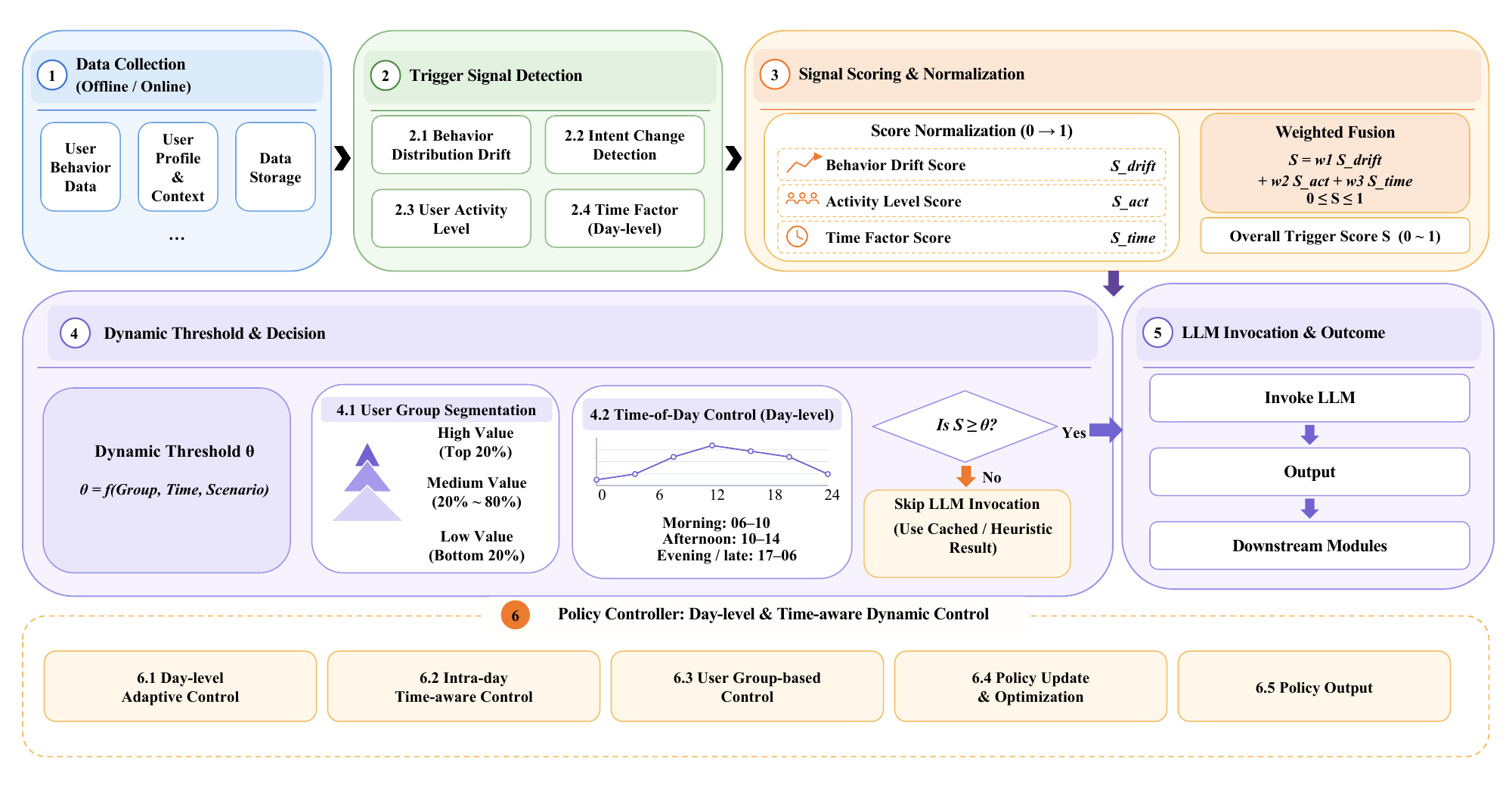}
    \caption{Intent-aware frequency control for LLM invocation. Behavioral drift, intent change, activity, and time signals are normalized and fused into an overall trigger score. A policy controller produces group-, time-, and scenario-dependent thresholds; only events exceeding the active threshold invoke the LLM, while other events use cached or heuristic results. Invocation outcomes are fed back to update the signal models and control policy.}
    \label{fig:intent_frequency_control}
\end{figure}

Intent inference with LLMs improves user modeling, but invoking an LLM after every event is wasteful because intent is often stable across adjacent interactions. We therefore formulate frequency control as a selective-computation problem: \textbf{when does the expected value of refreshing intent justify its cost?} As shown in \Cref{fig:intent_frequency_control}, the policy detects behavioral change, fuses normalized trigger signals, and applies context-dependent thresholds.

\subsubsection{Problem Formulation}
\label{sec:frequency_problem}

Let $u\in\mathcal U$ denote a user; let W denote the length of the recent-history window; let $\tau$ denote the event epoch within this window; and let $t$ denote an event-driven decision epoch. The recent history is $\mathcal H_{u,t}^{W}=\{e_{u,\tau}\}_{\tau=t-W+1}^{t}$, where events include \texttt{search}, \texttt{click}, \texttt{exposure}, \texttt{purchase}, and \texttt{post-purchase} actions. The decision context is defined as
\begin{equation}
    x_{u,t}=\big(\mathcal{H}_{u,t}^{W},\,p_u,\,c_{u,t},\,q_{u,t}^{\mathrm{cache}},\,h_{u,t}^{\mathrm{call}}\big),
\end{equation}
where $p_u$ is the stable profile, $c_{u,t}$ includes the current channel, device, scenario, and time, $q_{u,t}^{\mathrm{cache}}$ is the cached intent, and $h_{u,t}^{\mathrm{call}}$ is the recent invocation history.

The binary action $a_{u,t}\in\{0,1\}$ either invokes the LLM to update the intent summary and intent score or reuses a cached/heuristic result. Both paths feed the existing ranking pipeline. Given downstream utility $R(q;x)$, the value of a counterfactual refresh is defined as
\begin{equation}
    \Delta_{u,t}=
    \mathbb{E}\!\left[
        R(q_{u,t}^{\mathrm{llm}};x_{u,t})-
        R(q_{u,t}^{\mathrm{cache}};x_{u,t})
        \mid x_{u,t}
    \right].
    \label{eq:counterfactual_gain}
\end{equation}
Let $C_{u,t}^{\mathrm{llm}}$ and $L_{u,t}^{\mathrm{llm}}$ denote invocation cost and latency. For each user group $g$ and time bucket $b$, the constrained policy is formulated as
\begin{align}
    \max_{\pi}\quad
    &\mathbb{E}_{\pi}\!\left[
        a_{u,t}\Delta_{u,t}
        -\lambda_c a_{u,t}C_{u,t}^{\mathrm{llm}}
        -\lambda_l a_{u,t}L_{u,t}^{\mathrm{llm}}
    \right],
    \label{eq:frequency_objective}\\
    \text{s.t.}\quad
    &\mathbb{E}_{\pi}\!\left[
        \sum_{(u,t)\in\mathcal{B}_{g,b}} a_{u,t}
    \right]\leq B_{g,b},
    \qquad \forall(g,b),
    \label{eq:frequency_budget}
\end{align}
where $B_{g,b}$ is the invocation budget and $\lambda_c,\lambda_l\geq0$ are resource prices. Since $\Delta_{u,t}$ is unavailable online, the policy estimates refresh urgency through trigger signals.

Let $\widehat{\Delta}_{u,t}=\phi_{g,b,s}(S_{u,t})$ be a monotone calibration that maps the trigger score to the refresh value, and let $\mu_{g,b}\geq0$ be the dual price of capacity. The pointwise Lagrangian decision is
\begin{equation}
    a_{u,t}^{\star}=\mathds{1}\!\left[
        \widehat{\Delta}_{u,t}\geq
        \lambda_c C_{u,t}^{\mathrm{llm}}+
        \lambda_l L_{u,t}^{\mathrm{llm}}+
        \mu_{g,b}
    \right].
    \label{eq:lagrangian_decision}
\end{equation}
By monotonicity, \Cref{eq:lagrangian_decision} is equivalent to a context-dependent score threshold. Threshold adaptation therefore tracks both invocation value and the shadow price of LLM capacity.

\subsubsection{Intent-Aware Trigger Signal Modeling}
\label{sec:frequency_signals}

The detector derives four complementary signals from the feature store. A robust normalizer $\mathcal N_{j,g,s}$ maps each raw statistic to $[0,1]$ using group- and scenario-specific reference quantiles, preventing high-variance features from dominating solely because of their scale. $S_{u,t}\in[0,1]$ estimates how urgently the cached intent should be refreshed.

\paragraph{Behavior-distribution drift.}
Let $P_{u,t}^{\mathrm{cur}}$ and $P_{u,t}^{\mathrm{ref}}$ be current and historical feature distributions. Category, brand, action-type, and channel drift is measured by
\begin{equation}
    S_{u,t}^{\mathrm{drift}}=
    \mathcal N_{\mathrm{drift},g,s}\!\left(
    D_{\mathrm{JS}}(P_{u,t}^{\mathrm{cur}}\Vert P_{u,t}^{\mathrm{ref}})\right),
\end{equation}
where $D_{\mathrm{JS}}$ is the symmetric Jensen--Shannon divergence.

\paragraph{Activity level and time factor.}
Let $n^{\mathrm{pv}}$, $n^{\mathrm{uv}}$, and $n^{\mathrm{act}}$ denote Page View (PV), Unique Visitor (UV), and high-value action counts, respectively. Activity and intra-day propensity are
\begin{align}
    S_{u,t}^{\mathrm{act}}=
    \mathcal{N}_{\mathrm{act},g,s}\!\left(
        \log\!\left(1+n_{u,t}^{\mathrm{pv}}+\alpha n_{u,t}^{\mathrm{uv}}
        +\beta n_{u,t}^{\mathrm{act}}\right)
    \right),\\
    S_{u,t}^{\mathrm{time}}=
    \mathcal N_{\mathrm{time},g,s}\!\left(F_{\mathrm{time}}(b(t),d(t),s)\right),
\end{align}
where $F_{\mathrm{time}}$ means the time factor, $b(t)$ identifies morning peak, afternoon, evening peak, or late night and $d(t)$ is day-level context. The four signals are fused as
\begin{equation}
    S_{u,t}=
    w_1S_{u,t}^{\mathrm{drift}}+
    w_2S_{u,t}^{\mathrm{act}}+
    w_3S_{u,t}^{\mathrm{time}},
    \quad
    w_i\geq0,\quad \sum_{i=1}^{4}w_i=1.
    \label{eq:trigger_fusion}
\end{equation}

\subsubsection{Dynamic Thresholding and Online Execution}
\label{sec:frequency_decision}

A fixed threshold cannot capture user value, traffic, scenario importance, and capacity simultaneously. We therefore decompose the threshold into a global base $\theta_0$ plus four additive corrections:

\begin{itemize}
    \item $\delta_{g(u)}$: user-group offset. Users are partitioned into groups $g(u)$ by historical value (e.g.\ high/medium/low), and each group receives a learned offset.
    \item $\delta_{b(t)}$: time-bucket offset. The day is divided into buckets $b(t)$ (e.g.\ peak vs.\ off-peak); this term raises $\theta$ under heavy load to suppress low-value calls.
    \item $\delta_{s(t)}$: scenario offset. Different request scenarios $s(t)$ (e.g.\ homepage refresh, search, detail page) carry distinct importance levels.
    \item $\delta_{d(t)}$: day-level drift correction, capturing slow traffic shifts across calendar days $d(t)$.
\end{itemize}

The final threshold is
\begin{equation}
    \theta_{u,t}=\operatorname{clip}\!\left(
        \theta_0+\delta_{g(u)}+\delta_{b(t)}+\delta_{s(t)}+\delta_{d(t)},
        \theta_{\min},\theta_{\max}
    \right).
    \label{eq:dynamic_threshold}
\end{equation}
This realizes $\theta=f(\text{Group},\text{Time},\text{Scenario})$ with a day-level correction. The figure illustrates thresholds $0.45$, $0.60$, and $0.75$ for high-, medium-, and low-value users. An eligibility mask $m_{u,t}\in\{0,1\}$ enforces cooldowns, per-user caps, scenario allowlists, and overload protection. The deployed decision is
\begin{equation}
    a_{u,t}=m_{u,t}\,\mathds{1}\!\left[S_{u,t}\geq\theta_{u,t}\right].
    \label{eq:invoke_decision}
\end{equation}
An invocation updates the intent cache and downstream features; otherwise the system follows the cached or heuristic path.

\subsection{Strategy Orchestration and Parameter Translation}
\label{sec:control_tools_orchestration}

\begin{figure}[htbp]
    \centering
    \includegraphics[width=\linewidth]{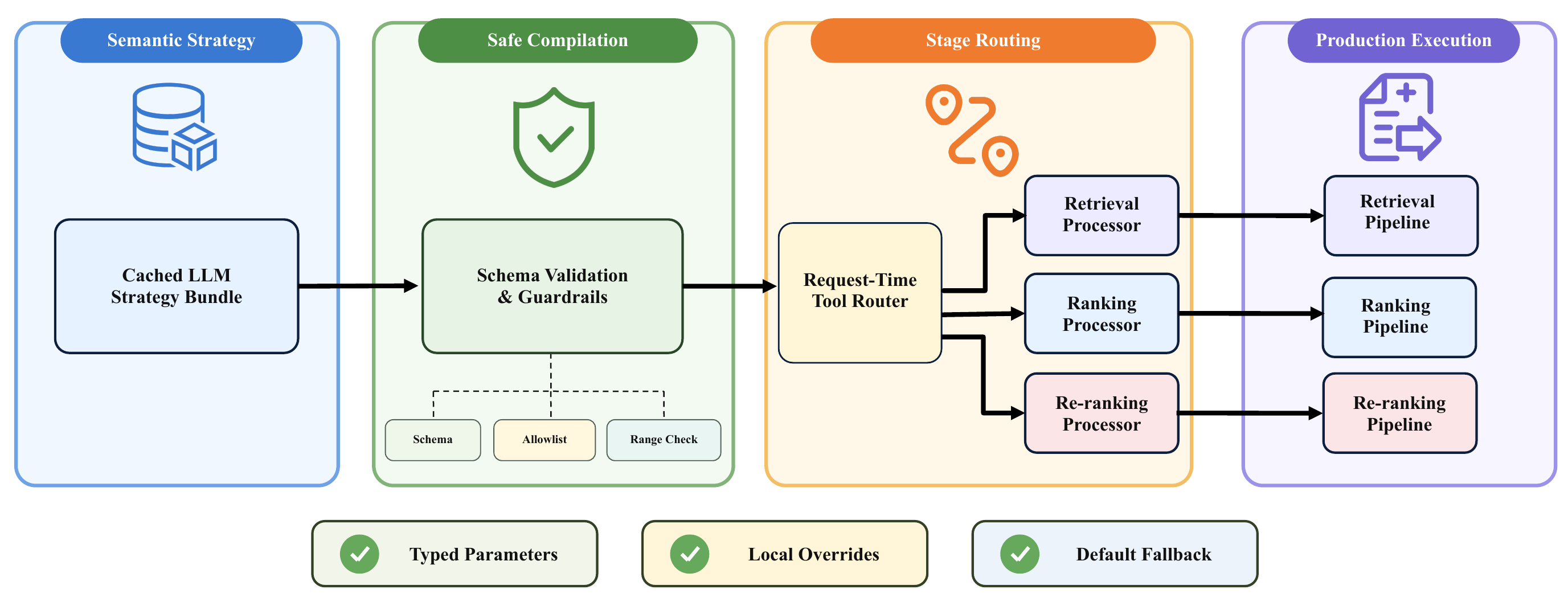}
    \caption{Request-time strategy orchestration. DREAM stores one semantic
    strategy bundle and compiles it into stage-specific, guarded parameter
    overrides. The production pipeline itself remains unchanged.}
    \label{fig:tools_orchestration}
\end{figure}

The MetaModel reasons in a semantic action space, whereas production services
consume typed numerical parameters, identifiers, filters, and quotas. DREAM
bridges these interfaces with a shared strategy contract and a set of
stage-specific Tool Processors. The contract is shared across the pipeline;
the parameter interpretation is local to each stage. This separation lets the
LLM coordinate coupled decisions without generating arbitrary service
configuration or replacing the existing retrieval, ranking, and re-ranking
implementations.

\subsubsection{Hierarchical Strategy Contract}

DREAM organizes a decision into three layers. M1 summarizes the current user
state and establishes a global orientation, including the decision rationale,
objective orientation, purchasing-power group, and activity group. M2 is the
schema-constrained JSON bundle generated by the MetaModel. It contains semantic
actions such as \texttt{objective boosts}, \texttt{business support}, \texttt{category} and \texttt{content-type
preferences}, \texttt{experience constraints}, and \texttt{position policies}. M3 is not
free-form LLM output. It is the deterministic result of compiling the M2
actions into parameters consumed by production services.

Table~\ref{tab:dream-strategy-contract} summarizes the current contract. A
field may be consumed by more than one stage, but each processor assigns it a
stage-appropriate operational meaning. For example, a \texttt{category preference} can
increase retrieval quota, relax a category scatter window during ranking, and
adjust category composition during re-ranking. The semantic preference is
shared; the underlying parameter is not.

\begin{table}[htbp]
    \centering
    \small
    \caption{Semantic M2 actions and their current stage-specific consumers.}
    \label{tab:dream-strategy-contract}
    \begin{tabularx}{\linewidth}{p{0.25\linewidth}p{0.34\linewidth}X}
        \toprule
        \textbf{M2 field} & \textbf{Semantic decision} &
        \textbf{Current consumer} \\
        \midrule
        \texttt{intent\_summary} & Summary of current intents and available
        supply & MetaModel planning context \\
        \texttt{ranking\_weight\_boost} & Relative emphasis on CTR, IPV, CVR,
        and GMV & Ranking and re-ranking \\
        \texttt{business\_support} & Whitelisted PlanIDs to protect & Ranking
        truncation and downstream mixing \\
        \texttt{category\_preference} & Preference over eligible categories or
        category-tag values & Retrieval, ranking, and re-ranking \\
        \texttt{cardtype\_preference} & Preference over product, video, live,
        and content supplies & Retrieval and re-ranking \\
        \texttt{experience\_constraints} & Exposure, purchase, and density
        constraints & Re-ranking \\
        \texttt{top\_ctr\_strategy} & Page-specific top-CTR and reverse-order
        switches & Re-ranking \\
        \bottomrule
    \end{tabularx}
\end{table}

\subsubsection{Request-Time Tool Orchestration}

LLM inference is decoupled from the latency-critical serving path. The
nearline MetaModel writes the latest valid strategy bundle to an online cache.
When a recommendation request arrives, the Tool Process reads the cached
bundle, validates its version and schema, and dispatches the relevant fields
to the retrieval, ranking, and re-ranking processors in
Figure~\ref{fig:tools_orchestration}. Let $a_u$ be the cached strategy for user
$u$, $p_s^0$ the default parameters of stage $s$, and $\mathcal{T}_s$ its
deterministic translator. The request-time configuration is defined as
\begin{equation}
    p_s^{\mathrm{exec}}=
    \operatorname{Guard}_s\!\left(
    p_s^0\oplus\mathcal{T}_s(a_u)\right),
    \quad
    s\in\{\mathrm{ret},\mathrm{rank},\mathrm{rerank}\},
    \label{eq:dream-tool-process}
\end{equation}
where $\oplus$ denotes a local override rather than a replacement of the full
configuration. A processor reads only fields on its allowlist. Missing,
expired, or malformed bundles therefore leave the corresponding default
parameters unchanged.

\subsubsection{Stage-Specific Parameter Translation}

\paragraph{Retrieval.}
The retrieval processor currently uses category and content-type preferences.
For \texttt{category\_preference}, it extracts the selected category-tag
values, serializes them into the cache field
\texttt{metamodel\_boost\_cate}, and uses the
\texttt{metamodel-boost} queue to protect retrieval volume for those
categories. For \texttt{cardtype\_preference}, the current implementation
supports a hard video-supply suppression action: when the semantic value for
video is $-2$, it emits
\path{metamodel_adjust_cardtype=video}, instructing the retrieval service
not to recall video content. Other content-type levels remain no-ops until a
corresponding whitelisted operator is implemented.

\paragraph{Ranking.}
The ranking processor consumes three M2 modules. First,
\texttt{ranking\_weight\_boost} modifies the original learning-to-rank score
without replacing it. The supported objectives are CTR, IPV, CVR, and GMV.
For each objective $i$, the processor reads a configured base weight $w_i^0$
and a bounded semantic level $b_i\in\{-2,-1,0,1,2\}$, then computes
\begin{equation}
    \delta_i=w_i^0b_i.
    \label{eq:dream-ltr-delta}
\end{equation}
The ranking service consumes
\texttt{ltr\_ctr\_delta}, \texttt{ltr\_ipv\_delta},
\texttt{ltr\_cvr\_delta}, and \texttt{ltr\_gmv\_} \texttt{delta} as multiplicative
corrections to the calibrated objective terms:
\begin{equation}
    f_{\mathrm{rank}}=
    f_{\mathrm{ltr}}
    \prod_i\left(1+\delta_i\widehat v_i\right),
    \label{eq:dream-ranking-score}
\end{equation}
where $\widehat v_i$ denotes the production-calibrated prediction used by the
corresponding objective.

Second, \texttt{business\_support.plan\_ids} is translated into
\texttt{guaranteed\_plan\_ids}. Items associated with these plans may bypass
the normal truncation quota, up to the configured
\path{business_support_max_num}. Third,
\texttt{category\_preference} controls candidate diversity before truncation.
For category $c$, a semantic ratio $b_c\in\{-2,-1,0,1,2\}$ becomes a scatter
window adjustment:
\begin{equation}
    n_c^{\mathrm{new}}=
    \max\!\left(1,n_c^{\mathrm{default}}+b_c\right).
    \label{eq:dream-scatter-window}
\end{equation}
Negative values enforce stronger dispersion; positive values permit greater
concentration when user intent is clear.

\paragraph{Re-ranking.}
The re-ranking processor applies the remaining joint strategy to the final
list construction. It combines objective-weight adjustments, category and
content-type preferences, exposure and purchase filters, density constraints,
and page-specific top-position rules. These actions configure one existing
production Generator or a whitelisted re-ranking recipe; the LLM does not emit
item identifiers or a final permutation.

\subsubsection{Validation, Isolation, and Fallback}

Every translation passes four gates: JSON parsing, schema validation,
stage-level allowlist checking, and parameter-range validation. Business
support is capped, scatter windows have a positive lower bound, and semantic
levels outside their enumerated domain are rejected. The compiled result is
scoped to the designated request and processor, so an override cannot mutate
unrelated stages or global configuration. When any gate fails, DREAM records
the failure for monitoring and executes the production default. This
default-plus-local-override design gives the MetaModel a useful control surface while preserving auditability, rollback, and the safety boundary of the existing pipeline.

\subsection{Offline RL Training Framework}
\label{sec:control_offline_rl}

\begin{figure}[htbp]
    \centering
    \includegraphics[width=\linewidth]{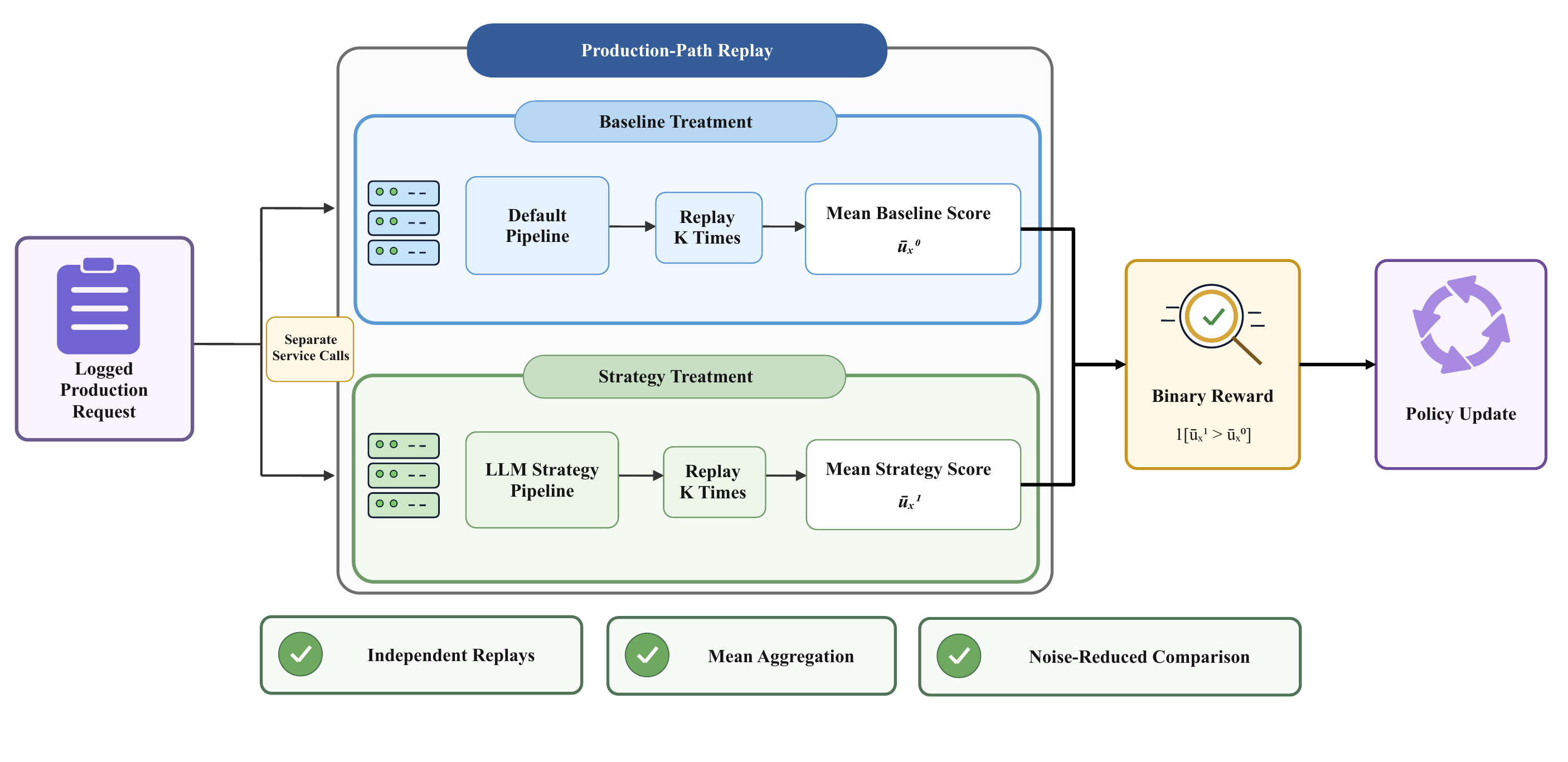}
    \caption{Offline policy training with production-path replay. The baseline
    and strategy treatments are separate service calls. Repeated execution and
    mean aggregation reduce serving noise before binary reward assignment.}
    \label{fig:offline_rl}
\end{figure}

DREAM trains its strategy policy offline while executing rollouts through the
production recommendation path. This design avoids exploratory decisions on
user-facing traffic and evaluates a strategy only after it has been translated
by the same Tool Processors used in serving. The learning problem is a
single-step contextual decision: given a logged request, the MetaModel produces
one strategy bundle, the production pipeline executes it, and a list-level
Evaluator supplies a proxy outcome for the final sequence. We do not simulate
subsequent clicks, transactions, or user-state transitions.

\subsubsection{Replay Dataset and Production Execution}

The replay dataset is constructed from input logs of real online requests. A
record retains the user and request context needed by the Intent Engine and
MetaModel, together with the identifiers and configuration required to invoke
the production service. For each record $x$, the environment constructs two
treatments:
\begin{align}
    L_{x,k}^{0} &= F(x;p^0,\xi_{x,k}^{0}),
    \label{eq:dream-baseline-replay}\\
    L_{x,k}^{1} &=
    F\!\left(x;
      p^0\oplus\mathcal{T}(a_x),\xi_{x,k}^{1}\right),
    \qquad a_x\sim\pi_\theta(\cdot\mid x),
    \label{eq:dream-strategy-replay}
\end{align}
where $F$ is the existing "retrieval--ranking--re-ranking" pipeline, $p^0$ is
its default configuration, and $\mathcal{T}$ denotes the collection of
stage-specific translators in Section~\ref{sec:control_tools_orchestration}.
The random variable $\xi$ represents the live feature state, model state, and
other serving variation encountered during an individual replay call. The baseline
treatment does not apply a MetaModel override; the strategy treatment reads
and executes the generated strategy through the request-time Tool Process.

The two treatments are sent as separate calls to the production service. Even
when they originate from the same log record, their outputs need not be
identical across repetitions because online features and production models are
resolved at replay time. We therefore execute each treatment $K$ times and
score every final list with the production list-level Evaluator $E$:
\begin{equation}
    u_{x,k}^{z}=E(L_{x,k}^{z},x),
    \qquad z\in\{0,1\},\quad k=1,\ldots,K.
    \label{eq:dream-replay-score}
\end{equation}
The environment aggregates repeated scores before comparing the two
treatments:
\begin{equation}
    \bar u_x^{z}=\frac{1}{K}\sum_{k=1}^{K}u_{x,k}^{z}.
    \label{eq:dream-mean-score}
\end{equation}
Mean aggregation reduces variance from independent service calls. It does not
turn replay into an exact reconstruction of the historical serving state, so
all conclusions remain relative to the production environment used during the
replay window.

\subsubsection{Binary Evaluator Reward}

DREAM uses a deliberately simple decision-aligned reward. A generated strategy
receives positive reward only when its mean Evaluator score exceeds the mean
score of the default pipeline for the same logged request:
\begin{equation}
    r(x,a_x)=
    \mathds{1}\!\left[\bar u_x^{1}>\bar u_x^{0}\right].
    \label{eq:dream-binary-reward}
\end{equation}
This comparison accounts for differences in the scale of Evaluator scores across requests and directly asks whether the personalized full-pipeline override is preferable to the production default. The policy objective is:
\begin{equation}
    \max_\theta\;
    J(\theta)=
    \mathbb{E}_{x\sim\mathcal{D},\,
    a_x\sim\pi_\theta(\cdot\mid x)}
    \left[r(x,a_x)\right].
    \label{eq:dream-rl-objective}
\end{equation}
The corresponding policy update increases the likelihood of executable
strategy bundles that win the repeated baseline comparison. Evaluator score
differences are retained as diagnostics, but the training reward itself is
binary; it does not combine rank, lift, curriculum competition, or teacher
distillation terms.

The Evaluator is a learned proxy for immediate list quality. Delayed online
outcomes such as clicks and transactions are not synthesized as episode
feedback in this offline loop. They belong to DREAM's online reward loop,
where observed outcomes can be used to monitor the policy, recalibrate the
Evaluator, and consolidate validated conclusions into Strategy Memory.

\subsubsection{Safe Replay and Pre-deployment Validation}

Replay calls use isolated load-test traffic. Replay calls return their output
lists to the training environment without exposing them to users or writing
impression, attribution, or business-counting records. The same schema,
translation logic, range checks, and fallback behavior are used during
training and serving. If a strategy is invalid, the cache misses, a
translator fails, or a parameter is out of range, the system falls back to
$p^0$ and records the event as an invalid policy action.

Before online rollout, we evaluate the policy on held-out request logs. The
primary offline measurements are binary win rate against the default pipeline,
mean Evaluator-score difference, strategy validity, Tool Process success rate,
and fallback rate. We additionally inspect results by user-state and intent
segments to ensure that aggregate wins do not hide systematic regressions.
Only policies that pass these checks proceed to guarded traffic ramp-up and
randomized online A/B testing. Because baseline and strategy treatments are
separate service calls, repeated averaging reduces but cannot eliminate
time-varying serving noise; randomized online evaluation remains the final
evidence of user and business impact.

\section{Experiments}
\label{sec:experiments}

We evaluate DREAM through production A/B tests and an end-to-end qualitative
case study. The main online experiment is a cumulative stage-wise ablation that
extends Meta Engine control from re-ranking to fine ranking, while the downstream
experiments evaluate how Intent Engine outputs improve individual production
applications. All online results are expressed as relative lifts over their
corresponding production baselines.

\subsection{Online Evaluation}
\label{sec:end_to_end_online_evaluation}
\label{sec:online_ab}

\paragraph{Setup and Metrics.}
\label{sec:online_metrics}
We conduct a production A/B test on Taobao Homepage Guess You Like. The
production baseline uses the incumbent pipeline. The two cumulative treatment
configurations enable the same Intent Engine, MetaModel, parameter translation,
and production safety guardrails, first adding re-ranking control and then fine-ranking control. This cumulative design isolates the conditional gain from
expanding control to the next production stage. We use the following online
metrics:
\begin{itemize}[leftmargin=*,itemsep=1pt,topsep=3pt]
    \item \textbf{PV} is the number of valid recommendation exposure events.
    \item \textbf{IPV} (Item Page Views) counts attributed item-detail-page
    visits.
    \item \textbf{Core IPV} restricts IPV to the core item scope.
    \item \textbf{PCTR} is the click-through rate after excluding
    back-navigation re-exposures.
    \item \textbf{Click UV} is the approximate number of distinct visitors
    with at least one valid one-hop click (\(\mathrm{dpv1}>0\)).
    \item \textbf{UCTR} is the ratio of Click UV to Exposure UV.
    \item \textbf{GMV} is the Gross Merchandise Value of the attributed paid
    orders.
    \item \textbf{Ad Cost} measures advertising spend and is treated as a
    positive measure of commercial value for the platform.
\end{itemize}
For each metric \(M\), we report the relative
lift \(\Delta M=(M_{\mathrm{treatment}}/M_{\mathrm{baseline}}-1)\times100\%\);
absolute control and treatment values are omitted for confidentiality.

\paragraph{Results and Discussion.}
\begin{table}[H]
\centering
\small
\caption{Cumulative stage-wise online lifts of Meta Engine over the production baseline.}
\label{tab:online_main}
\begin{threeparttable}
\resizebox{\linewidth}{!}{%
\begin{tabular}{lcccccccc}
\toprule
\textbf{Configuration}
& \textbf{PV \(\uparrow\)}
& \textbf{IPV \(\uparrow\)}
& \makecell{\textbf{Core}\\\textbf{IPV \(\uparrow\)}}
& \textbf{PCTR \(\uparrow\)}
& \textbf{Click UV \(\uparrow\)}
& \textbf{UCTR \(\uparrow\)}
& \textbf{GMV \(\uparrow\)}
& \makecell{\textbf{Ad}\\\textbf{Cost \(\uparrow\)}} \\
\midrule
DREAM (@ rerank)
& +1.03\%
& +2.06\%
& +2.39\%
& +0.76\%
& +0.54\%
& +0.53\%
& +0.88\%
& \textbf{+0.21\%}
\\
DREAM (@ rerank \& rank)
& \textbf{+1.04\%}
& \textbf{+2.71\%}
& \textbf{+3.06\%}
& \textbf{+1.25\%}
& \textbf{+0.68\%}
& \textbf{+0.81\%}
& \textbf{+1.31\%}
& +0.02\%
\\
\bottomrule
\end{tabular}%
}
\end{threeparttable}
\end{table}

Table~\ref{tab:online_main} reports the stage-wise online ablation as Meta Engine expands its control coverage. Re-ranking control alone improves all reported metrics over the production
baseline, including PV by \(1.03\%\), IPV by \(2.06\%\), Core IPV by \(2.39\%\), GMV by \(0.88\%\), and PCTR by \(0.76\%\). Extending Meta Engine to fine
ranking further raises IPV and Core IPV to \(+2.71\%\) and \(+3.06\%\), and GMV
to \(+1.31\%\), corresponding to additional gains of \(0.65\), \(0.67\), and
\(0.43\) percentage points over re-ranking alone; PCTR also increases from
\(+0.76\%\) to \(+1.25\%\). Click UV and UCTR
similarly rise from \(+0.54\%\) and \(+0.53\%\) to \(+0.68\%\) and \(+0.81\%\).
Meanwhile, PV remains nearly unchanged between the two treatments
(\(+1.03\%\) versus \(+1.04\%\)), indicating that the deeper-stage gains come
primarily from improved engagement and conversion of existing exposure rather
than additional exposure volume.

\subsection{Intent Engine Downstream Evaluation}
\label{sec:intent_downstream_ab}

\paragraph{Setup and Metrics.}
We further evaluate the Intent Engine through independent online A/B tests
on the downstream applications described in
Section~\ref{sec:intent_supply}.
In each test, the treatment integrates the corresponding intent fields into
the production scenario, and the baseline is the same scenario without intent signals. The results are reported at two levels of granularity: platform-wide effects for
applications measured across all recommendation streams
(Table~\ref{tab:online_intent_platform}) and in-scenario effects within
Heuristic Recommendation (Table~\ref{tab:online_intent_scenario}).
We use the following metrics at both levels:
\begin{itemize}[leftmargin=*,itemsep=1pt,topsep=3pt]
    \item \textbf{PV}, \textbf{IPV}, \textbf{Core IPV}, and \textbf{PCTR}
    follow the definitions in Section~\ref{sec:online_metrics}.
    \item \textbf{Inquiry Card Clicks} is the number of valid clicks on inquiry cards in the recommendation feed.
    \item \textbf{CTR} is the ratio of valid one-hop clicks—clicks on product items within the landing page triggered by an Inquiry Card—to the corresponding PV within the evaluated scenario.
    \item \textbf{Transaction Volume} is the number of completed orders attributed to the evaluated scenario.

\end{itemize}

\paragraph{Results and Discussion.}
\begin{table}[htbp]
\centering
\small
\caption{Platform-wide online A/B results for Intent Engine downstream
applications.}
\label{tab:online_intent_platform}
\begin{threeparttable}
\begin{tabular*}{\linewidth}{@{\extracolsep{\fill}}lcccc@{}}
\toprule
\textbf{Application}
& \textbf{IPV \(\uparrow\)}
& \textbf{Core IPV \(\uparrow\)}
& \textbf{PCTR \(\uparrow\)}
& \textbf{Transaction Volume \(\uparrow\)} \\

\midrule
Recall (Cognitive Recommendation)
& +0.80\% & +0.91\% & +0.84\% & +0.36\% \\
Recommendation Strategy Adaptation
& +0.52\% & +0.70\% & -0.02\% & +0.33\% \\

\bottomrule
\end{tabular*}
\end{threeparttable}
\end{table}

\begin{table}[htbp]
\centering
\small
\caption{In-scenario online A/B results within Heuristic Recommendation.}
\label{tab:online_intent_scenario}
\begin{threeparttable}
\begin{tabular*}{\linewidth}{@{\extracolsep{\fill}}lccc@{}}
\toprule
\textbf{Application}
& \textbf{PV \(\uparrow\)}
& \textbf{Inquiry Card Clicks \(\uparrow\)}
& \textbf{CTR \(\uparrow\)} \\
\midrule
Recall (Heuristic Recommendation)
& +5.06\% & +7.17\% & +2.00\% \\
Explicit Text Interaction
& +7.41\% & +10.64\% & +3.01\% \\
\bottomrule
\end{tabular*}
\end{threeparttable}
\end{table}

At the platform level (Table~\ref{tab:online_intent_platform}), both
applications deliver consistent gains. All reported metrics show positive
lifts. Intent-driven recall in Cognitive Recommendation improves
platform-wide Core IPV by \(0.91\%\), IPV by \(0.80\%\), and PCTR by
\(0.84\%\). Strategy adaptation yields a \(0.70\%\) lift in Core IPV and
\(0.52\%\) in IPV while maintaining a neutral PCTR (\(-0.02\%\)),
indicating that the strategy adjustment improves conversion depth without
sacrificing click-through rate. Both applications also contribute to
the growth of the transaction volume (\(+0.36\%\) and \(+0.33\%\)). Given that
platform-wide metrics are considerably more difficult to improve than
in-scenario metrics, these lifts are meaningful in practice.
Within Heuristic Recommendation (Table~\ref{tab:online_intent_scenario}), the two optimizations are complementary: recall integration determines which needs are surfaced as inquiry cards, while copy optimization determines how each need is verbalized to the user. These two optimizations lift inquiry card clicks by \(7.17\%\) and \(10.64\%\) respectively, and both improve
CTR (\(+2.00\%\) and \(+3.01\%\)).

\subsection{Case Study}
\label{sec:case_study}

\begin{figure}[H]
    \centering
    \includegraphics[width=\linewidth]{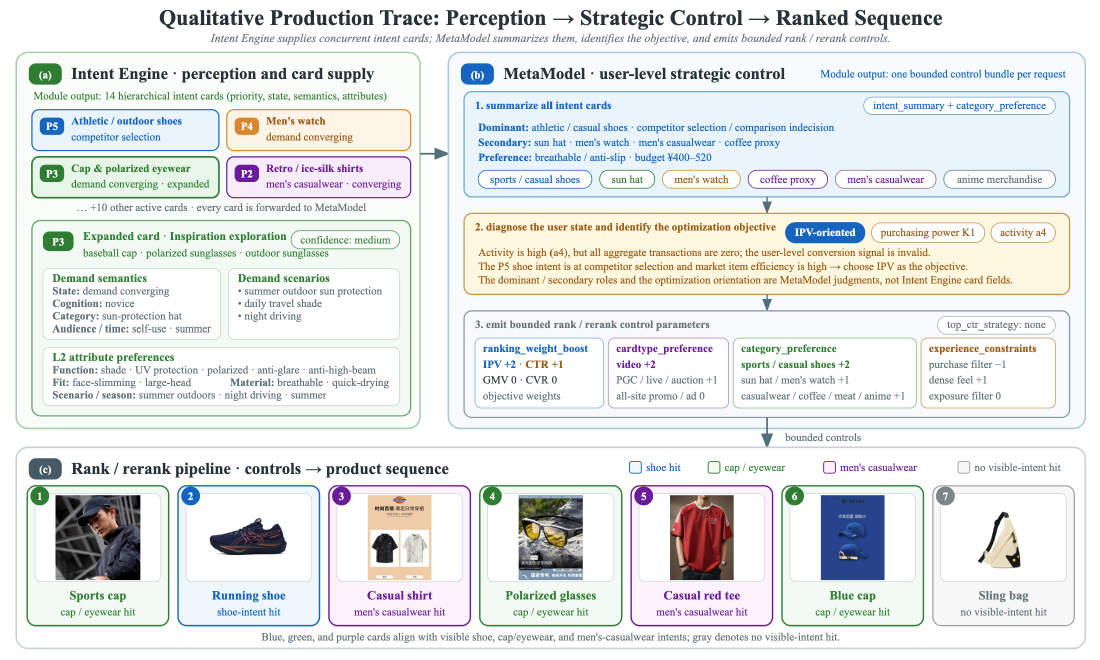}
    \caption{Qualitative production trace through the DREAM control path.
    (a)~The Intent Engine supplies multiple concurrent hierarchical intent
    cards; P3 is expanded only to expose the card schema. (b)~MetaModel jointly summarizes the cards, diagnoses the user state, identifies the optimization objective, and emits bounded rank/re-rank control parameters. (c)~The rank/re-rank pipeline applies the selected overrides and returns the product sequence.}
    \label{fig:case_study_intent_trace}
\end{figure}

Figure~\ref{fig:case_study_intent_trace} traces one anonymized production
request across the module boundaries of DREAM while retaining the returned
product creatives. It separates the Intent Engine's structured perception
output, MetaModel's strategic reasoning and control bundle, and execution by
the rank/re-rank pipeline.

The Intent Engine emits multiple concurrent cards rather than one merged
intent. Panel~(a) shows representative P5, P4, P3, and P2 cards from the 14 active
intents and expands one P3 card only to expose its schema. Priority, decision
state, demand semantics, confidence, and L2 attributes belong to the Intent
Engine output. The expanded card describes inspiration exploration anchored by
baseball caps and polarized or outdoor sunglasses. Its demand is converging
with medium confidence, and it retains novice category cognition,
sun-protection-hat demand, summer and night-driving scenarios, and functional,
fit, and material preferences.
The P2 card captures converging novice demand for men's casualwear across
commuting, casual, summer, and outdoor use, with retro, ice-silk, tie-dye,
breathable, loose/workwear, and black/red/printed preferences.

MetaModel begins only after receiving the complete card set. In its first
stage, it emits \texttt{intent\_\allowbreak summary} and
\texttt{category\_\allowbreak preference}: it identifies athletic/casual shoes
as the dominant intent at competitor selection or comparison indecision, with
sun-protection hats, men's watches, men's casualwear, and coffee proxy ordering
as secondary intents,
and summarizes breathable or anti-slip preferences and a CNY~400-520 budget.
The dominant/secondary assignment is therefore a MetaModel judgment rather
than an Intent Engine field.

MetaModel next combines that summary with the user strata. The request has
purchasing-power level \(K1\) and activity level \(a4\), but all aggregate
transaction-behavior counts are zero, so the user-level conversion signal is treated
as invalid. Because the P5 shoe intent is already at competitor selection and
market-level item efficiency is high, MetaModel selects an IPV-oriented
objective. It then parameterizes that decision as a sparse control bundle:
\texttt{IPV}=+2 and \texttt{CTR}=+1 while \texttt{GMV}=\texttt{CVR}=0;
video receives +2; selected card types and categories receive +1 or +2; and
experience constraints receive bounded adjustments such as purchase filtering
at -1 and dense-layout preference at +1.

Panel~(c) shows the rank/re-rank pipeline consuming the bounded overrides and
returning the product sequence. Blue cards match the visible shoe intent,
green cards match the visible cap/eyewear intent, purple cards match the P2
men's-casualwear intent, and gray cards have no match to the intent cards
displayed in the figure. This color coding denotes semantic alignment, not
causal attribution to any single control parameter. Additional Intent Engine
cases appear in Appendix~\ref{sec:intent_engine_cases}.

\section{Conclusion}
\label{sec:conclusion}

We have presented DREAM, an autonomous optimization control architecture that adds a perception-aware, orchestrable, and auditable policy layer atop conventional retrieval--ranking--re-ranking pipelines without replacing any existing model. Its two core components---an Intent Engine that fuses hundreds of on-device signals into structured L0/L1/L2 intent representations, and a Meta Engine that translates intent into executable strategy bundles via layered reasoning augmented by Strategy Memory---are continuously optimized by a Reward Dual Loop that couples offline simulation with online feedback for continuous self-improvement, together closing the long-standing gap between intent perception, strategy generation, and execution optimization. Large-scale A/B testing on Taobao's Homepage Feed validates the architecture: with re-ranking control alone, DREAM achieves \hlnum{+2.06\%} IPV, \hlnum{+2.39\%} Core IPV, and \hlnum{+0.88\%} GMV; extending control to fine ranking further raises these to \hlnum{+2.71\%} IPV, \hlnum{+3.06\%} Core IPV, and \hlnum{+1.31\%} GMV, with PV consistently increasing by more than \hlnum{1\%}. The cumulative gains from deeper pipeline integration confirm that the benefits of agentic meta-control compound as the control surface expands, thereby validating agentic meta-control as a practical paradigm for industrial recommendation optimization.

\addcontentsline{toc}{section}{References}
\bibliographystyle{abbrvnat}
\nobibliography*
\bibliography{reference}

\clearpage

\appendix
\section*{Appendix}
\section{Contributors}

\begin{multicols}{2}
\noindent
\textcolor[HTML]{5b0f08}{
\textbf{Core Contributors} \\
Bin Zhang \\
Bowen Zheng \\
Chao Yi \\
Chengyu Lai \\
Dian Chen \\
Dimin Wang \\
Gaoyang Guo \\
Jialin Zhu \\
Jian Wu \\
Jing Yu \\
Jiuning Lin \\
Lingqing Zhang\textsuperscript{$\dagger$} \\
Lingyun Zheng \\
Mao Zhang \\
Mingming Pan \\
Ruiquan Lan\textsuperscript{$\dagger$} \\
Shuai Zhong\textsuperscript{$\dagger$} \\
Wen Chen \\
Wendong Zhang \\
Xiaodong Zhu\textsuperscript{$\dagger$} \\
Xuan Chen\textsuperscript{$\dagger$} \\
Xunke Xi \\
Yifan Lu\textsuperscript{$\dagger$} \\
Yiheng Wang\textsuperscript{$\dagger$} \\
Yue Zeng\textsuperscript{$\dagger$} \\
Yujie Luo \\
Yuning Jiang \\
Zhe Hu \\
Zhibo Xiao \\
Zihong Huang
}


\noindent
\textcolor[HTML]{030361}{
\textbf{Contributors} \\
Binbin Cao \\
Bo Zheng \\
Danning Wang \\
Dixuan Wang \\
Ge Fan \\
Haixia Wu \\
Han Zhu \\
Hao Fang \\
Haoming Chen \\
Huiping Chu \\
Jian Wang \\
Jianjun Wu \\
Jiawei Wu \\
Jiaxin Yu \\
Jingwen Liu\textsuperscript{$\dagger$} \\
Jinzhe Shan \\
Kai Meng \\
Kai Zhang \\
Keqin Xu \\
Kewei Zhu \\
Lang Tian \\
Leihui Chen \\
Li Chen \\
Licheng Xu \\
Lide Xiao \\
Ruitong Zhang \\
Shiyao Peng\textsuperscript{$\dagger$} \\
Silu Zhou \\
Tao Wang \\
Wei Shi \\
Wenjun Yang \\
Xiang Chen \\
Xiang Gao \\
Xiao Ren \\
Xu Liu \\
Xuwen Wang \\
Yang Li \\
Yeqiu Yang \\
Yi Hu \\
Yichen Yuan \\
Yinnan Song \\
Yipeng Yu \\
Yuan Liu \\
Yunqi Gao \\
Zhiliang Huang \\
Zhujin Gao \\
Zongyuan Wu
}
\end{multicols}

\noindent
The authors are listed alphabetically by first name.

\noindent
$\dagger$ Work done during a summer internship at Taobao \& Tmall Group of Alibaba.

\newpage

\definecolor{interest_colframe}{rgb}{0.8, 0.878, 0.871}
\definecolor{interest_colback}{rgb}{0.918, 0.953, 0.949}

\definecolor{tag_colframe}{rgb}{0.965, 0.898, 0.847}
\definecolor{tag_colback}{rgb}{0.988, 0.961, 0.941}

\definecolor{exp_colback}{rgb}{0.949, 0.965, 0.980}
\definecolor{exp_colframe}{rgb}{0.878, 0.922, 0.965}

\tcbset{
    promptbox/.code args={#1/#2}{
        \tcbset{
            enhanced,
            arc=0mm,
            colframe=#1, 
            colback=#2, 
            coltitle=black,
            fonttitle=\large\bfseries,
            attach boxed title to top left={xshift=0mm, yshift=-1.0mm},
            boxed title style={
                skin=enhancedfirst jigsaw,
                size=small,
                arc=3mm,
                bottom=0mm,
                left=8mm,
                right=8mm,
                top=1mm,
                colback=#1
            },
            boxrule=0pt,
            frame hidden,
            borderline north={4pt}{0pt}{#1},
        }
    }
}

\section{Implementation Details}
\label{sec:impl_details}

\subsection{Policy Execution}
\label{sec:metamodel_exec}

\begin{table}[htbp]
\centering
\caption{Controllable parameters of the policy execution layer.  L1 selects a sparse strategy subset per request; L2 maps it to concrete values within guardrail-enforced safe ranges.}
\label{tab:controllable_params}
\small
\begin{tabularx}{\textwidth}{@{}llL@{}}
\toprule
\textbf{Stage} & \textbf{Module} & \textbf{Control Mechanism} \\
\midrule
\multirow{3}{*}{Retrieval}
  & \texttt{BE Chain}           & Toggle retrieval paths on/off. \\
  & \texttt{Plan / Category Quota} & Reallocate quota across plans and categories. \\
  & \texttt{Ranking Formula}    & Adjust CTR / IPV / CVR / GMV term weights. \\
\midrule
\multirow{4}{*}{Blending}
  & \texttt{Natural Blending Alpha}   & Set per-objective weights in the blending formula. \\
  & \texttt{Global-Push Alpha Hat}    & Scale alpha multiplier for non-global-push buckets. \\
  & \texttt{Extreme-CTR Strategy}     & Switch slot-level formula between CTR-aggressive and balanced. \\
  & \texttt{PVR Control Weighting}    & Set the coefficient used to merge the ad and boost control scores. \\
\midrule
Boosting
  & \texttt{Boost Weighting}          & Inject weighted boost scores for designated items/plans. \\
\midrule
\multirow{6}{*}{Experience}
  & \texttt{Category Diversity}       & Set minimum same-category card scatter interval. \\
  & \texttt{CSR Filtering}            & Remove cards below per-card-type satisfaction threshold. \\
  & \texttt{DTR Filtering}            & Remove cards above per-card-type dislike threshold. \\
  & \texttt{Personalized Fatigue}     & Adjust fatigue decay coefficient. \\
  & \texttt{Vector Scatter}           & Configure embedding-similarity spacing threshold. \\
  & \texttt{Rule-based Scatter}       & Set spacing by card type / brand / seller / category. \\
\midrule
Re-ranking
  & \texttt{Generator}                & Inject additional generator modules. \\
\bottomrule
\end{tabularx}
\end{table}

The execution layer delivers Meta Engine decisions to the online pipeline without replacing any existing module.  It uses a design that combines fallback to the default configuration with overrides based on personalized instructions: the baseline configuration keeps the pipeline stable, while the M3 parameter translation module selectively overrides a subset of parameters per request, all bounded by safety guardrails.  Table~\ref{tab:controllable_params} lists the parameters currently exposed, organized by pipeline stage.

The parameters span both relevance-oriented knobs (retrieval paths, ranking weights, blending alpha) and experience-oriented knobs (diversity, fatigue, filtering).  The MetaModel does not manipulate all of them simultaneously; given the current user intent from the Intent Engine, M2 selects a sparse subset aligned with the strategy orientation, and M3 translates it into concrete values.  Because every override is incremental and bounded by the guardrails, the mainline pipeline continues to function as the safety net, and DREAM's interventions remain auditable and reversible.

\subsection{Offline MetaModel Policy Evaluation}
\label{sec:offline_metamodel_eval}

\paragraph{Setup and Metrics.}
We evaluate the 4B MetaModel through production-path replay on Taobao Homepage
Guess You Like. The Base and RL policies use the same held-out requests,
strategy prompt, typed action schema, validator, compiler, execution engine,
and production Evaluator. The comparison therefore isolates the contribution
of replay RL while holding model scale and the execution path fixed. We report:
\begin{itemize}[leftmargin=*,itemsep=1pt,topsep=3pt]
    \item \textbf{pCTR}, the pointwise click-through-rate diagnostic;
    \item \textbf{pCVR}, the pointwise conversion-rate diagnostic;
    \item \textbf{pIPV}, the pointwise item-page-view diagnostic;
    \item \textbf{pGMV}, the pointwise GMV diagnostic;
    \item \textbf{Valid}, the fraction of generated strategy bundles that pass
    schema validation and compile into executable strategies.
\end{itemize}
Base serves as the reference, and the RL row reports
\((M_{\mathrm{RL}}/M_{\mathrm{Base}}-1)\times100\%\) for each metric \(M\).

\paragraph{Results and Discussion.}
\begin{table}[H]
\centering
\small
\caption{Offline relative lifts of the 4B replay-RL policy over its Base.}
\label{tab:offline_rl}
\begin{threeparttable}
\begin{tabular*}{\linewidth}{@{\extracolsep{\fill}}lccccc@{}}
\toprule
\textbf{Policy}
& \textbf{pCTR \(\uparrow\)}
& \textbf{pCVR \(\uparrow\)}
& \textbf{pIPV \(\uparrow\)}
& \textbf{pGMV \(\uparrow\)}
& \textbf{Valid \(\uparrow\)} \\
\midrule
RL (4B) & \textbf{+2.42\%} & -0.99\% & \textbf{+1.38\%}
& \textbf{+0.37\%} & \textbf{+22.25\%} \\
\bottomrule
\end{tabular*}
\end{threeparttable}
\end{table}

Relative to the 4B Base, replay RL improves pCTR by \(2.42\%\), pIPV by
\(1.38\%\), and pGMV by \(0.37\%\), while pCVR changes by \(-0.99\%\).
Validity improves by \(22.25\%\) in relative terms, corresponding to an
increase from \(80.86\%\) to \(98.85\%\), or \(17.99\) percentage points.
Overall, replay RL substantially improves strategy executability while
producing positive changes on three of the four pointwise diagnostics.

\section{Prompt Formats}
\label{app:prompt-formats}

\subsection{Intent Engine Prompt}
\label{app:intent-engine-prompt}

\begin{tcolorbox}[promptbox=tag_colframe/tag_colback, title={Intent Engine Prompt -- Hierarchical Intent Reasoning},label=box:intent-engine-prompt,breakable]

\colorbox{green!20}{\strut\textbf{\# Role}}\\[3pt]
You are the core intent-understanding engine of an e-commerce intent platform. Given modular user signals, infer structured user intents at two levels: \textbf{L1 Demand Layer} and \textbf{L2 Preference Layer}. You should reason like an expert in user psychology, consumer behavior and product-domain knowledge. Every judgment must be supported by explicit signals; do not hallucinate unsupported preferences.\\[6pt]

\colorbox{blue!20}{\strut\textbf{\# Input Modules}}\\[3pt]
All input modules are optional. Reason only from the modules actually provided; do not make assumptions about missing modules.\\[3pt]
\textbf{User Data}: static L0 attributes, long-term behavior across months or seasons, and short-term browsing/clicking/cart/purchase signals.\\[2pt]
\textbf{On-device Behavior}: real-time session signals such as dwell time, scroll direction and speed, clicked targets, search queries, detail-page comparison, review inspection, add-to-cart and checkout actions.\\[2pt]
\textbf{Historical Intent}: the previous intent structure or a compressed summary of multiple rounds. If present, fuse it with the newest signals through continuation, correction or overwrite; never copy it mechanically.\\[6pt]

\colorbox{red!20}{\strut\textbf{\# Reasoning Rules}}\\[3pt]
\textbf{Signal priority}. Real-time behavior $>$ short-term behavior $>$ long-term behavior $>$ static attributes. Fresher behavioral evidence has higher weight.\\[2pt]
\textbf{Historical-intent fusion}. If new signals agree with a historical intent, strengthen confidence and refine preferences. If they conflict, follow the new evidence and down-weight the historical intent. If evidence is insufficient to overturn it, keep the historical intent but lower confidence. If an intent is unchanged in the current turn, do not output an \texttt{update}.\\[2pt]
\textbf{Intent type}. Use \textit{goal-driven} when there is a clear product/category target and purchase motivation; use \textit{inspiration exploration} when the user browses broadly within a loosely defined area without strong purchase intent; use \textit{aimless browsing} when the user's behavior is random and non-convergent, with short dwell times and no search, cart, or favorite signals.\\[2pt]
\textbf{Decision state}. Choose from: demand vague, demand converging, competitor comparison, price deliberation, decision bottleneck, about-to-buy, post-purchase exit, post-purchase verification, active abandonment and demand suspension. Once a purchase has been made in a category, the corresponding intent state can move only to post-purchase exit or post-purchase verification, not back to a pre-purchase state.\\[2pt]
\textbf{Real-time psychology}. Infer only from behavior patterns when possible; otherwise output \texttt{null}. Candidate states include price deliberation, attribute conflict, information anxiety, cognitive focus, trust verification, content fatigue, aimless wandering, expectation gap, substitute hesitation and impulse rollback.\\[2pt]
\textbf{Uncertainty}. When evidence is insufficient, prefer \texttt{null} to fabricated attributes. The output should remain predictive of the user's likely future clicks, searches or purchases.\\[6pt]

\colorbox{orange!20}{\strut\textbf{\# Multi-intent Handling}}\\[3pt]
Split independent intents when any of the following is present: unrelated category clusters; different target audiences; nonoverlapping usage scenarios; or a clear switch in the behavior sequence, such as an unrelated query jump.\\[2pt]
Merge behaviors into the same intent when categories are complementary, belong to the same shopping task, or the category switch reflects comparison, outfit/usage matching or basket-building rather than an intent jump.\\[2pt]
Assign a \texttt{priority} from 1 to 5. Stronger real-time evidence ranks above short-term, long-term and static evidence; later decision stages rank higher; ties are broken by recency.\\[6pt]

\colorbox{purple!18}{\strut\textbf{\# Field Constraints}}\\[3pt]
1. Unknown scalar fields must be \texttt{null}; unknown list fields must be empty arrays.\\
2. Multiple values in \texttt{demand category} or \texttt{refined category} must be separated by commas (,).\\
3. \texttt{Refined category} must be concrete, searchable, and not over-fragmented.\\
4. Attribute preferences must follow \texttt{attribute:value1,value2}.\\
5. For updates, list fields must be emitted as complete replacement arrays rather than element-level diffs.\\[6pt]

\colorbox{cyan!18}{\strut\textbf{\# Routing Self-Assessment}}\\[3pt]
Always produce the synchronous \texttt{actions} first. Then assess whether this request is too complex or uncertain for the $0.8$B Main Agent and should be asynchronously refined by a larger model. This routing decision does not block the current output; it only tells the router whether to enqueue a correction task.\\[2pt]
Set a high \texttt{route\_confidence} when any of the following holds: the behavior sequence is too long or information-dense; one request spans many unrelated L1 categories; the current L1 category is plausible but the L2 refined category or schema update requires longer memory; the L1 category is fixed but concrete item fields such as brand tendency, price preference, attributes, decision state or real-time psychology remain uncertain; or the request involves rare, high-value or cross-category reasoning that may exceed the Main Agent's capacity.\\[2pt]
Choose \texttt{route.tool} as follows: \texttt{none} when no asynchronous refinement is needed; \texttt{context\_subagent} when intent structure or category reasoning requires broader behavioral context; and \texttt{expert} when the category is established but item-level preference fields require further refinement. Use \texttt{route\_confidence=null} only when no explicit model-level escalation signal is produced and the rule layer should decide alone.\\[6pt]

\colorbox{yellow!25}{\strut\textbf{\# Output Protocol}}\\[3pt]
Output strict JSON only. Do not output explanations outside JSON. The top-level object is \texttt{\{"actions": [...], "route": \{...\}\}}.\\[3pt]
\textbf{Operation types}:\\
\texttt{insert}: use when detecting new intents. The \texttt{intent} field is a list of complete intent objects.\\
\texttt{update}: use when an existing intent changes. Match by \texttt{intent\_id}; changed fields are represented by dot-path keys.\\[2pt]
\textbf{Route fields}: \texttt{tool} is one of \texttt{none}, \texttt{context\_subagent}, \texttt{expert}; \texttt{route\_confidence} is a scalar in $[0,1]$ or \texttt{null}. Higher values indicate a stronger need for asynchronous escalation.\\[4pt]

\fcolorbox{brown!20}{yellow!5}{\parbox{0.96\linewidth}{\ttfamily\small\raggedright
\{\\
\hspace*{1em}"actions": [\\
\hspace*{2em}\{"op": "update", "id": "<intent\_id>", "fields": \{"<dot.path>": "<new\_value>"\}\},\\
\hspace*{2em}\{"op": "insert", "intent": ["<complete intent object with metadata, L1, L2>"]\}\\
\hspace*{1em}],\\
\hspace*{1em}"route": \{\\
\hspace*{2em}"tool": "<none | context\_subagent | expert >",\\
\hspace*{2em}"route\_confidence": <float in [0,1] | null>\\
\hspace*{1em}\}\\
\}
}}

\end{tcolorbox}

\subsection{Meta Engine Prompt}
\label{app:meta-engine-prompt}

\begin{tcolorbox}[promptbox=interest_colframe/interest_colback, title={Prompt Template (Meta Engine -- Strategy Orchestration)},label=box:meta-engine-prompt,breakable]

\colorbox{green!20}{\strut\textbf{\# Role}}\;\;You are \texttt{MetaBrain}, the meta-model that orchestrates the homepage feed recommendation pipeline. All downstream components execute the strategy bundle you output.\\[6pt]

\colorbox{blue!20}{\strut\textbf{\# Input}}\;\;Each turn provides two payload segments:\\[3pt]
\textbf{env\_state} --- Post-hoc feed metrics (1/3/7-day windows). Share and efficiency metrics for six card types (\texttt{product / ad / universal\_push / short\_video / live / content}) are provided at both the market and user levels. Trend: \texttt{a>b>c} denotes a rising trend, whereas \texttt{a<b<c} denotes a declining trend. If user-level data are sparse, fall back to market-level data.\\[3pt]
\textbf{user\_intent} --- Structured intent JSON with three tiers:\\
\quad\textbf{L0}\;Stable profile: \texttt{activity} $\in$ \{a1..a5\}, \texttt{purchasing\_power} $\in$ \{K1..K6\}, demographics.\\
\quad\textbf{L1}\;Per-intent: \texttt{demand\_category}, \texttt{confidence} $\in$ \{low, med, high\}, \texttt{intent\_type} $\in$ \{goal-driven, inspiration exploration, aimless browsing\}.\\
\quad\textbf{L2}\;Fine-grained: \texttt{brand\_tendency}, \texttt{price\_preference}, \texttt{realtime\_psychology}, \texttt{decision\_state} (one of the ten decision states, from \textit{demand vague} to \textit{demand suspension}).\\[6pt]

\colorbox{red!18}{\strut\textbf{\# Output}}\;\;A strategy bundle: 1 M1 module + 6 M2 modules (fixed order). All values must be selected from the specified enumerations.\\[4pt]

\textbf{M1 --- Overall Baseline}\\
\quad \texttt{decision\_rationale}: a chain-of-thought anchor ($\leq$100 characters); it must be the first key.\\
\quad \texttt{strategy\_orientation} $\in$ \{\texttt{IPV-oriented}, \texttt{GMV-oriented}\};\quad \texttt{purchasing\_power} $\in$ \{K1..K6\};\quad \texttt{activity} $\in$ \{a1..a5\}.\\[4pt]

\textbf{M2 --- Six Modules.} \textit{Level semantics}: \texttt{+2} = strongest boost / \texttt{+1} = mild boost / \texttt{0} = neutral / \texttt{-1} = mild suppression / \texttt{-2} = strongest suppression.\\[4pt]

\fcolorbox{violet!30}{violet!3}{\parbox{0.95\linewidth}{\strut\textbf{1.\;intent\_summary}\;{\small (intent-forwarding)}\\[2pt]
\texttt{intent\_summary}: str $\leq$100 chars, dominant intent summary.\quad \texttt{category\_preference}: list of demand categories for retrieval.}}\\[4pt]

\fcolorbox{violet!38}{violet!6}{\parbox{0.95\linewidth}{\strut\textbf{2.\;ranking\_weight\_boost}\;{\small (objective-preference)}\\[2pt]
\texttt{ctr / ipv / cvr / gmv\_boost\_ratio}, each $\in$\{-2..2\}. Adjusts the $\alpha$ weights in the scoring formula: $\text{score}=\text{ctr}\cdot(\alpha_\text{ctr}+\alpha_\text{ipv}\!\cdot\!\text{ipv}+\alpha_\text{cvr}\!\cdot\!\text{cvr}+\alpha_\text{gmv}\!\cdot\!\text{gmv})$. Only level \textit{differences} matter.}}\\[4pt]

\fcolorbox{violet!46}{violet!9}{\parbox{0.95\linewidth}{\strut\textbf{3.\;cardtype\_preference}\;{\small (attribute-preference)}\\[2pt]
\texttt{auction / ad / quanzhantui / video / live / pgc}, each $\in$\{-2..2\}. All six keys are required; at least three must be non-zero, including at least one negative value.}}\\[4pt]

\fcolorbox{violet!54}{violet!12}{\parbox{0.95\linewidth}{\strut\textbf{4.\;category\_preference}\;{\small (attribute-preference)}\\[2pt]
\texttt{"<real\_category\_name>": int}, each $\in$\{-2..2\}. Use real category names; the total number of keys must not exceed six.}}\\[4pt]

\fcolorbox{violet!62}{violet!15}{\parbox{0.95\linewidth}{\strut\textbf{5.\;experience\_constraints}\;{\small (attribute-preference;\;\texttt{+}=tighten, \texttt{-}=relax)}\\[2pt]
\texttt{exposure\_filter} (dedup exposed items)\;/\;\texttt{purchase\_filter} (filter purchased items)\;/\;\texttt{density\_perception} (similar-content diversity), each $\in$\{-2..2\}.}}\\[4pt]

\fcolorbox{violet!70}{violet!18}{\parbox{0.95\linewidth}{\strut\textbf{6.\;top\_ctr\_strategy}\;{\small (strategy-control; binary switch)}\\[2pt]
\texttt{page0 / pagen} $\in$\{0,1\}. Enables pure-CTR ranking on target slots. \textbf{Overrides} the $\alpha$-weighted formula from module~2.}}\\[6pt]

\colorbox{orange!20}{\strut\textbf{\# Key Decision Principles}}\\[3pt]
Every module fuses \textbf{intent-driven signals} (\texttt{user\_intent}) with \textbf{post-hoc signals} (\texttt{env\_state}); user-level signals are preferred, with market-level signals used as a fallback.\\[2pt]
$\bullet$ \textbf{Orientation}: early-stage intents $\Rightarrow$ IPV-oriented;\; late-stage (about-to-buy, decision-stuck) $\Rightarrow$ GMV-oriented.\\
$\bullet$ \textbf{Boost alignment}: IPV-oriented boosts \{ctr, ipv\}, suppresses \{cvr, gmv\}; GMV-oriented reverses. No signal $\Rightarrow$ \texttt{\{0,0,0,0\}}.\\
$\bullet$ \textbf{Cardtype pipeline}: trend protection $\rightarrow$ market comparison $\rightarrow$ user calibration $\rightarrow$ dominant constraint $\rightarrow$ intent calibration (env $>$ intent).\\
$\bullet$ \textbf{CTR switch}: only for first-click-rate bottlenecks; GMV-oriented + late decision states $\Rightarrow$ forced off.\\[6pt]

\colorbox{yellow!25}{\strut\textbf{\# Output Format}}\;\;Return a single JSON object inside \texttt{<answer>\{...\}</answer>}. Two keys: \texttt{M1} (object) + \texttt{M2} (list of 6 objects, each with \texttt{name / type / config}).\\[4pt]
\fcolorbox{brown!20}{yellow!5}{\parbox{0.95\linewidth}{\ttfamily\small\raggedright\strut
<answer>\{"M1": \{"decision\_rationale": "...",\\
\hspace*{2em}"strategy\_orientation": "IPV-oriented",\\
\hspace*{2em}"purchasing\_power": "K3", "activity": "a3"\},\\
"M2": [\\
\hspace*{1em}\{"name":"intent\_summary", ...\},\\
\hspace*{1em}\{"name":"ranking\_weight\_boost",\\
\hspace*{2em}"config":\{"ctr":0, "ipv":2, "cvr":-1, "gmv":-1\}\},\\
\hspace*{1em}\{"name":"cardtype\_preference",\\
\hspace*{2em}"config":\{"auction":1, "ad":-1, "quanzhantui":0,\\
\hspace*{4em}"video":2, "live":-1, "pgc":1\}\},\\
\hspace*{1em}\{"name":"category\_preference",\\
\hspace*{2em}"config":\{"dresses":2, "sneakers":1\}\},\\
\hspace*{1em}\{"name":"experience\_constraints",\\
\hspace*{2em}"config":\{"exposure\_filter":0,\\
\hspace*{4em}"purchase\_filter":0, "density\_perception":0\}\},\\
\hspace*{1em}\{"name":"top\_ctr\_strategy",\\
\hspace*{2em}"config":\{"page0":0, "pagen":0\}\}\\
]\}</answer>
}}
\end{tcolorbox}

\section{Intent Engine Cases}
\label{sec:intent_engine_cases}
The cases in this appendix are constructed illustrative examples that reflect typical user-modeling patterns in production. All attribute values, behaviors, and brand placeholders (e.g., Brand~A) are representative composites and do not correspond to any real user.

\definecolor{l0_bg}{RGB}{252, 248, 245}
\definecolor{l0_frame}{RGB}{191, 97, 106}
\definecolor{original_box}{RGB}{255, 250, 240}
\definecolor{compressed_box}{RGB}{240, 248, 255}
\definecolor{action_color}{RGB}{163, 73, 164}

\refstepcounter{caseexample}\label{case:l0_physical_layer}%
\noindent
\begin{tcolorbox}[
enhanced,
colback=l0_bg,
colframe=l0_frame,
boxrule=2pt,
arc=4mm,
width=\textwidth,
title={\textbf{Case \thecaseexample: L0 Physical Layer Memory Representation}},
coltitle=white,
fonttitle=\bfseries,
attach boxed title to top left={yshift=-3mm, xshift=6mm},
boxed title style={colback=l0_frame, arc=3mm},
breakable
]
\vspace{0.1cm}
\textcolor{blue!70!black}{\textbf{User Profile \& Behavioral Logs}}:

\vspace{0.2cm}
\colorbox{original_box}{
\begin{minipage}{0.95\linewidth}
\textbf{Basic Profile:}\\
Age band: 40--45 | Gender: female | Location: a coastal city in East China\\
Device: high-end smartphone | Purchasing power: relatively high |
Skin type: non-sensitive | Home-renovation demand: low

\vspace{0.2cm}
\textbf{Chronic Interests:}\\
Tea culture, collecting fine tableware, outdoor sports gear, children's
education, building blocks and model collecting, stockpiling household cleaning
supplies, health and wellness

\vspace{0.2cm}
\textbf{Behavioral Memory Snippet:}\\
\textcolor{action_color}{Early 2024 -- late 2025: home renovation and
decoration} | Prefers New Chinese--style interiors with solid-wood
materials \ldots recently focused on ambient-lighting upgrades and wall
maintenance \ldots attentive to decorative details (entryway paintings /
festive ornaments) \ldots reflecting an emphasis on festive rituals and
needs related to moving house | Brands: several furniture, lighting, and paint brands\ldots

\vspace{0.2cm}
\textbf{Identity Tags:}\\
core decision-maker for household purchases; mother of multiple children; currently in a period of peak education spending

\end{minipage}
}
\end{tcolorbox}

\definecolor{l1_bg}{RGB}{245, 250, 255}
\definecolor{l1_frame}{RGB}{52, 152, 219}

\refstepcounter{caseexample}\label{case:l1_demand_layer}%
\noindent
\begin{tcolorbox}[
enhanced,
colback=l1_bg,
colframe=l1_frame,
boxrule=2pt,
arc=4mm,
width=\textwidth,
title={\textbf{Case \thecaseexample: L1 Demand Layer Memory Representation}},
coltitle=white,
fonttitle=\bfseries,
attach boxed title to top left={yshift=-3mm, xshift=6mm},
boxed title style={colback=l1_frame, arc=3mm},
breakable
]
\vspace{0.1cm}
\textcolor{blue!70!black}{\textbf{Meta-Intent \& Category Demand}}:

\vspace{0.2cm}
\colorbox{original_box}{%
\begin{minipage}{0.95\linewidth}
\textbf{Demand Category:}\\
Multi-functional food processor

\vspace{0.2cm}
\textbf{Category Cognition:}\\
Moderately informed---has extensively browsed multiple models from
mainstream Brands A, B, and C, focusing on stainless-steel build, large
capacity, and high power; still at the comparison stage

\vspace{0.2cm}
\textbf{Usage Scenarios:}\\
Daily use in the family kitchen for meat grinding, blending, and other food-preparation tasks

\vspace{0.2cm}
\textbf{Target Audience:}\\
Self-use; family cooking needs

\vspace{0.2cm}
\textbf{Temporal Context:}\\
Concentrated search and browsing in late February 2026; a newly emerged
demand

\vspace{0.2cm}
\textbf{Confidence Level:}\\
High: searched ``meat grinder'' repeatedly within a few days and clicked
seven products from different brands; the activity is temporally concentrated and
unambiguous

\vspace{0.2cm}
\textbf{Inference Rationale:}\\
After repeated searches, the user repeatedly clicked on multiple household
electric meat grinders from Brands A, B, and C, covering different
functions and capacities. We infer a clear purchase intent at the
brand-and-model comparison stage, with Brand A receiving the most
attention.

\end{minipage}%
}
\end{tcolorbox}

\definecolor{l2_bg}{RGB}{245, 250, 245}
\definecolor{l2_frame}{RGB}{76, 175, 80}

\refstepcounter{caseexample}\label{case:l2_preference_layer}%
\noindent
\begin{tcolorbox}[
enhanced,
colback=l2_bg,
colframe=l2_frame,
boxrule=2pt,
arc=4mm,
width=\textwidth,
title={\textbf{Case \thecaseexample: L2 Preference Layer Memory Representation}},
coltitle=white,
fonttitle=\bfseries,
attach boxed title to top left={yshift=-3mm, xshift=6mm},
boxed title style={colback=l2_frame, arc=3mm},
before skip=1.5em plus 0.5em,
breakable
]
\vspace{0.1cm}
\textcolor{blue!70!black}{\textbf{Fine-Grained Preference under Active Demand}}:

\vspace{0.2cm}
\begin{tcolorbox}[
colback=original_box,
colframe=original_box,
boxrule=0pt,
arc=0pt,
left=2pt, right=2pt, top=2pt, bottom=2pt,
width=0.95\linewidth
]
\textbf{Subcategory:}\\
Brand A large-capacity multi-functional food processor

\vspace{0.2cm}
\textbf{Brand Preference:}\\
Prefers mainstream appliance brands; the user showed the greatest interest in Brand A, clicking on 5--6
Brand A meat grinders. Brands B and C followed, and the
preference is clearly concentrated on Brand A.

\vspace{0.2cm}
\textbf{Attribute Preference:}\\
\textit{Preferred:}\\
\quad $\bullet$ Multi-functionality: blending and baby-food preparation
beyond meat grinding\\
\quad $\bullet$ Stainless-steel build: durable and food-safe\\
\quad $\bullet$ Large capacity and high power for family
use\\
\textit{Excluded:} none

\vspace{0.2cm}
\textbf{Decision State:}\\
Competitive-selection stage: the user has extensively browsed multiple products from Brands A, B, and C but has not yet added any product to the cart or researched any product in depth.

\vspace{0.2cm}
\textbf{Engine Directive:}\\
Spawn subagent: household meat-grinder shopping expert

\end{tcolorbox}
\end{tcolorbox}

\clearpage

\end{CJK*}
\end{document}